\documentclass[%
 aip,
 amsmath,amssymb,
 reprint,%
]{revtex4-1}

\usepackage{graphicx}% Include figure files
\usepackage{dcolumn}% Align table columns on decimal point
\usepackage{bm}% bold math
\usepackage{upgreek}
\usepackage{xcolor}

\usepackage[utf8]{inputenc}
\usepackage[T1]{fontenc}
\usepackage{mathptmx}
\usepackage{etoolbox}

\usepackage[pdftex,colorlinks=true,allcolors=blue,breaklinks=true]{hyperref}

\usepackage{parskip}
\makeatletter
\def\@email#1#2{%
 \endgroup
 \patchcmd{\titleblock@produce}
  {\frontmatter@RRAPformat}
  {\frontmatter@RRAPformat{\produce@RRAP{*#1\href{mailto:#2}{#2}}}\frontmatter@RRAPformat}
  {}{}
}%
\makeatother
\begin{document}

\preprint{AIP/123-QED}

\title[BLS spectroscopy with AR-HCFs]{Low Spontaneous Brillouin Scattering in Anti-Resonant Hollow-Core Fibers in GHz Frequency Range}

\author{Ryan E. Dunagin}
 \affiliation{Center for Magnetism and Magnetic Nanostructures, Department of Physics and Energy Science, University of Colorado Colorado Springs, Colorado Springs, CO 80918, USA}
 
\author{Robbie Mears}%
\affiliation{Centre for Photonics and Photonic Materials, Department of Physics, University of Bath, Bath BA2 7AY, UK}

\author{Dario Bueno-Baques}
 \affiliation{Center for Magnetism and Magnetic Nanostructures, Department of Physics and Energy Science, University of Colorado Colorado Springs, Colorado Springs, CO 80918, USA}
 
\author{Vasyl S. Tyberkevych}
 \affiliation{Department of Physics, Oakland University, Rochester, MI 48309, USA}%Lines break automatically or can be forced with \\

\author{Yi Li}
 \affiliation{Materials Science Division, Argonne National Laboratory, Lemont, IL 60439, USA}

\author{William J. Wadsworth}
\affiliation{Centre for Photonics and Photonic Materials, Department of Physics, University of Bath, Bath BA2 7AY, UK}

\author{Zbigniew Celinski}
 \affiliation{Center for Magnetism and Magnetic Nanostructures, Department of Physics and Energy Science, University of Colorado Colorado Springs, Colorado Springs, CO 80918, USA}

\author{Valentine Novosad}
 \affiliation{Materials Science Division, Argonne National Laboratory, Lemont, IL 60439, USA}

\author{Dmytro A. Bozhko}\email{dbozhko@uccs.edu}
 \affiliation{Center for Magnetism and Magnetic Nanostructures, Department of Physics and Energy Science, University of Colorado Colorado Springs, Colorado Springs, CO 80918, USA} 

\date{\today}% It is always \today, today,

\begin{abstract}
Brillouin light scattering (BLS) is a powerful experimental tool that can be used to get insights into the fundamental and applied properties of matter, like dispersions of quasiparticles in a solid, as well as their spatio-temporal dynamics. Many applications of light scattering favor the use of optical fibers in place of free-space optics. In this work, we compare the performance of anti-resonant hollow core fibers to that of conventional solid core fused silica fibers for BLS experiments in the GHz frequency range. Conventional fibers are barely suitable for low-noise measurements because of the spontaneous scattering of the photons on various phononic modes present in the core and cladding. In the case of the hollow-core fiber, we identify a range of discrete phononic modes and associate them with the various acoustic modes of the structure surrounding the hollow core using finite-element numerical simulations. The measured relative intensity of the spontaneous BLS signal from these modes is orders of magnitude smaller than that of a solid-core fiber, making anti-resonant hollow-core fibers one of the best solutions for the single-mode light guidance for BLS and potentially other low-noise photonic experiments.
\end{abstract}

\maketitle

\section{\label{sec:Intro}Introduction}

Brillouin light scattering (BLS) spectroscopy has emerged as a powerful and versatile tool for studying the elastic and magnetic properties of a broad range of materials \cite{Bozhko2020, Han2004, Sebastian2015, Vavassori2000, Booth1987, Zhang1993, Belmeguenai2017, Bottcher2021, Schneider2013, Yudistira2016, Ye2025}. Its extraordinary sensitivity down to thermal population levels and simultaneous access to frequency, time, space, phase, and wavevector domains makes it essential in fundamental research and practical applications in photonics, magnetism, and telecommunications \cite{Fohr2009, Vogt2009, Demidov2004, Demidov2015, Madami2011, Heussner2020, Dunagin2025, Hillebrands1989, Buettner2000, Sebastian2015microBLS, Xia1998, Zhang1993, Hahn2022, Kunz2024, Frey2021, OrdezRomero2009, Sandercock1973, Buchmeier2007, Wettling1975, CryerJenkins2025, Mizuno2010, Kosareva2022, Ji2024, Chauhan2021, Liu2021, Gundavarapu2018}. The quality and reliability of BLS measurements depend significantly on the performance of the optical components used in the setup. Typical BLS experiments as referenced above are being performed using free-space optics, but in some applications (e.g., in cryogenics \cite{Hgele2008, Christiansen2023, Tokizaki1999}), it could be much more convenient, or even necessary, to use optical fibers \cite{Mahar2008, Beugnot2014, Tan2006}. \looseness=-1

Conventional solid-core fused silica fibers, widely employed due to their well-developed manufacturing technologies and established optical characteristics, exhibit inherent limitations in applications. They can produce substantial spontaneous (as well as stimulated) photon scattering from various phononic modes within the silica core and cladding \cite{Beugnot2014}, increasing background noise in a broad frequency range and thus limiting the signal-to-noise ratio or even making measurements impossible. Due to these spurious effects, solid silica fibers could pose considerable challenges or even make impossible some low-noise applications like quantum measurements and communications \cite{Shirasaki1992, Bergman1991, Shelby1986, Bergman1993, Feng2017, Lassen2013, Li2014}.  \looseness=-1

Recently, anti-resonant hollow core fibers (AR-HCFs) \cite{Fokoua2023} have gained attention as a promising solution for low-loss and low-noise photonic applications, especially for UV light guidance \cite{Mears2024_guidance}. By confining the guiding mode primarily within a gas- or vacuum-filled core, AR-HCFs can significantly mitigate losses associated with the fiber material itself \cite{Fokoua2023}. Absence of core material can also greatly reduce the number of undesired scattering events and, therefore, improve signal-to-noise ratio in BLS applications (for example, stimulated BLS in MHz frequency range \cite{Iyer2020, Renninger2016, Zhong2015}). In this study, we compare the performance of AR-HCFs to conventional fused silica fiber in the context of spontaneous BLS spectroscopy at GHz frequencies. We identified the main channels of spontaneous BLS in AR-HCFs and demonstrated that they are related to discrete acoustic modes in the capillary microstructures with resonant frequencies determined mainly by the capillary wall thickness. Notably, the intensities of the corresponding BLS peaks are suppressed by more than three orders of magnitude compared to those observed in solid-core fibers. Our findings establish AR-HCFs as highly advantageous for single-mode guidance and low-noise BLS experiments, providing new opportunities to advance the field of fiber-based photonic sensing, metrology, and quantum communications.  \looseness=-1

\section{\label{sec:Methods}Experimental Setup and Methodology}
% FIGURE 1
\begin{figure}
\includegraphics[width=1\columnwidth]{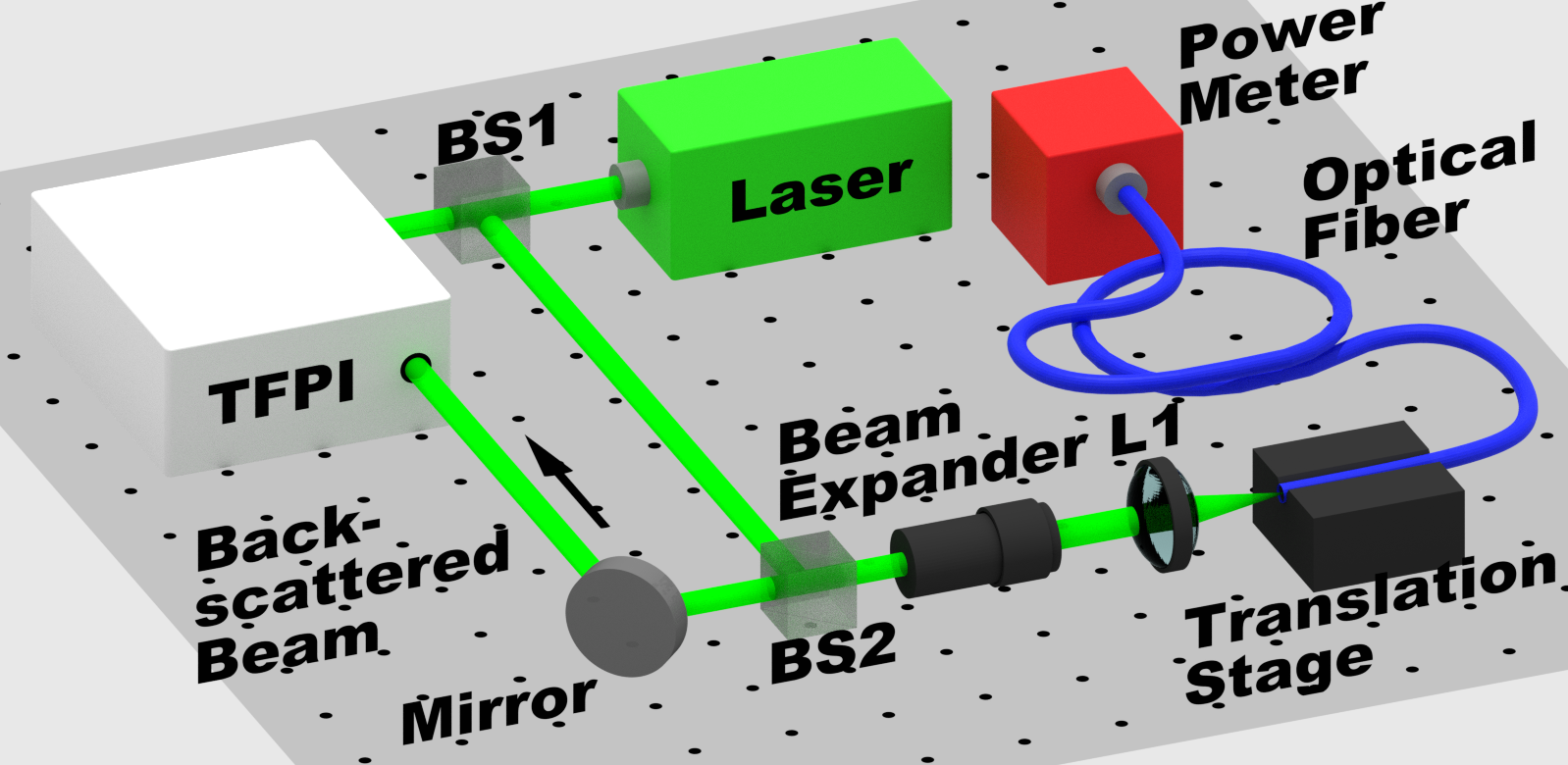}
\caption{\label{Fig1} Experimental setup for BLS spectroscopy of optical fibers. BS1 splits the beam into a reference beam and a probing beam. A beam expander is used to adjust the beam diameter incident onto the lens L1, used to couple the light into the optical fiber. The backscattered light is analyzed by frequency using Tandem Fabry-P\'{e}rot Interferometer (TFPI). The detailed description of the setup is in the main text of the article.}
\end{figure} 

The optical setup used for our study of the BLS method in optical fibers is shown in Fig.\,\ref{Fig1}. The excitation part of the setup consists of the following: a light source (which is a $532\,\mathrm{nm}$ wavelength single-mode single-frequency laser), guiding and beam-shaping optics, a fiber coupling lens L1 (which also serves as a backscattered light collecting element), and a multi-axis translation stage used for fiber precise alignment. The optical power meter is used to measure the transmission through the fiber to optimize its alignment and allow for BLS data normalization. The light scattered backwards from the fiber is collected by L1 and is guided to the Tandem Fabry-P\'{e}rot Interferometer (TFPI), where scattered photons are discriminated by frequency. The 6-pass geometry of the instrument allows for achieving extreme contrast of better than $10^{10}$ as well as obtaining frequency resolution down to $50\,\mathrm{MHz}$ \cite{Cardona2000}, which is essential for discrimination between different phononic modes observed in the fibers studied in this work. The beam sampled by the TFPI after BS1 is used for stabilization and frequency reference purposes.  \looseness=-1

It should be emphasized that in the case of spontaneous BLS, which is mostly used in studies of quasiparticle dynamics, the scattering occurs on thermally or otherwise forcedly excited states\cite{Cardona2000, Merklein2022}. In contrast, the stimulated BLS involves coherent scattering driven by modulations in the material’s dielectric constant induced by a strong optical field interference between incident and scattered light\cite{Chiao1964, Ballmann2015, Merklein2022, AlDabbagh2016, Corredera2013}. This process has a threshold, which scales with the fiber’s cross-sectional area and inversely with its length. In our measurements, we kept the optical power below $1\,\mathrm{mW}$ to ensure no emergence of stimulated effects in the studied fibers.

First, we performed measurements on the commercially available solid-core single-mode polarization-maintaining (PM) fiber (Thorlabs PM-S405-XP). The choice of this particular fiber was made based on its broad optical transmission band ($400-680\,\mathrm{nm}$) as well as the fact that in BLS spectroscopy it is very helpful to maintain and analyze the information about the scattered photons' polarization. For example, it is possible to discriminate between magnon quasiparticles, which rotate the plane of polarization by $90^{\circ}$\cite{Dunagin2025}, from longitudinal phonons by using a simple polarizer at the input of the TFPI. The measured photon counts were processed in the following way. The average dark noise counts were subtracted from the data, followed by normalization by the transmitted power and the number of TFPI scan passes performed to collect each spectrum. Transmitted power was used for normalization under the assumption that fibers have negligible losses, which is true for the $2\,\mathrm{m}$ long PM fiber used in the experiments reported here. This procedure allows for the comparison of the BLS efficiency for different fibers. 

\begin{figure}
\includegraphics[width=1\columnwidth]{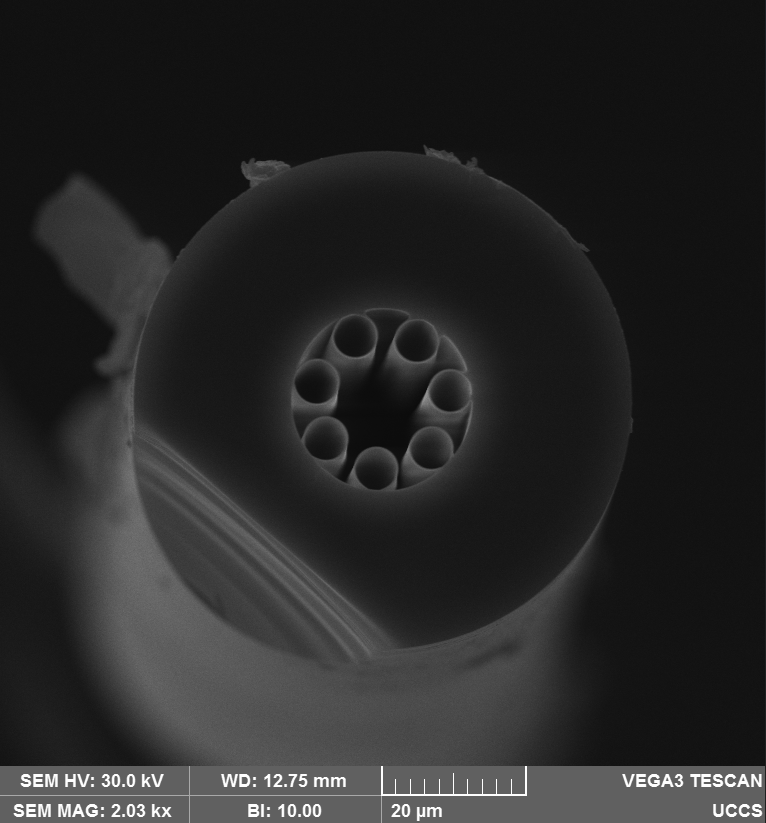}
\caption{\label{Fig2} Anti-resonant hollow core fiber cross-section imaged using SEM. The protective acrylate coating was removed, and a few nm of Au coating was applied before imaging to reduce charging. The thickness of seven antiresonance cylindrical capillary microstructures was determined by this technique to be $200\,\mathrm{nm}$. The inner diameter housing these microstructures is approximately $26\,\mathrm{\upmu m}$ and the diameter of the capillaries is $5.5\,\mathrm{\upmu m}$}
\end{figure}

Next, we performed the same measurement procedure on a $1.25\,\mathrm{m}$ long sample of AR-HCF. The particular fiber under test has been fabricated using a two-stage stack-and-draw method \cite{Murphy2022} out of low-OH synthetic silica (F300 Heraeus). The AR-HCF is additionally protected by a layer of acrylate coating. A Scanning Electron Microscope (SEM) image of the cross-section of the studied AR-HCF is shown in Fig.\,\ref{Fig2}. The light guidance is achieved by Fabry-P\'{e}rot resonators formed by thin walls (of about $200\,\mathrm{nm}$ thickness) of the silica capillaries surrounding the air core, resulting in leaky confinement of light within so-called ``anti-resonant'' wavelength ranges\cite{Litchinitser2002}. The resulting AR-HCF has a loss better than $0.5\,\mathrm{dB/m}$ across the entire transmission bands of $220-320\,\mathrm{nm}$ and $400-750\,\mathrm{nm}$\cite{Mears2024_guidance}. While AR-HCFs do not have a strict modal cutoff, the 7 capillary design ensures that the fiber is effectively single mode by resonant filtering of the higher-order modes \cite{Uebel2016}.  \looseness=-1

% FIGURE 2

To couple the probing laser light into each fiber, we used lenses matched to the numerical aperture (NA) and a translation and tip-tilt stage. For the conventional solid-core fiber with a NA of $0.12$, we used a Thorlabs CFC11P-A adjustable collimator. For the AR-HCF with a NA of $<0.04$ we used a $75\,\mathrm{mm}$ focal distance achromatic lens. The small NA of the AR-HCF meant that additional care needed to be taken to ensure the beam's parallelism to the fiber's axis. Provided that the probing beam was close to the optical axis of the collimating lens L1 (see Fig.\,\ref{Fig1}), small translations of the beam resulted in a very fine adjustment of the incident angle at the aperture of the fiber. After optimizing throughput, the transmitted power was measured after the optical fiber and was used (along with the fiber's length) for the BLS data normalization.  \looseness=-1

\section{\label{sec:Results}Results and Discussion}

\begin{figure}
\includegraphics[width=1\columnwidth]{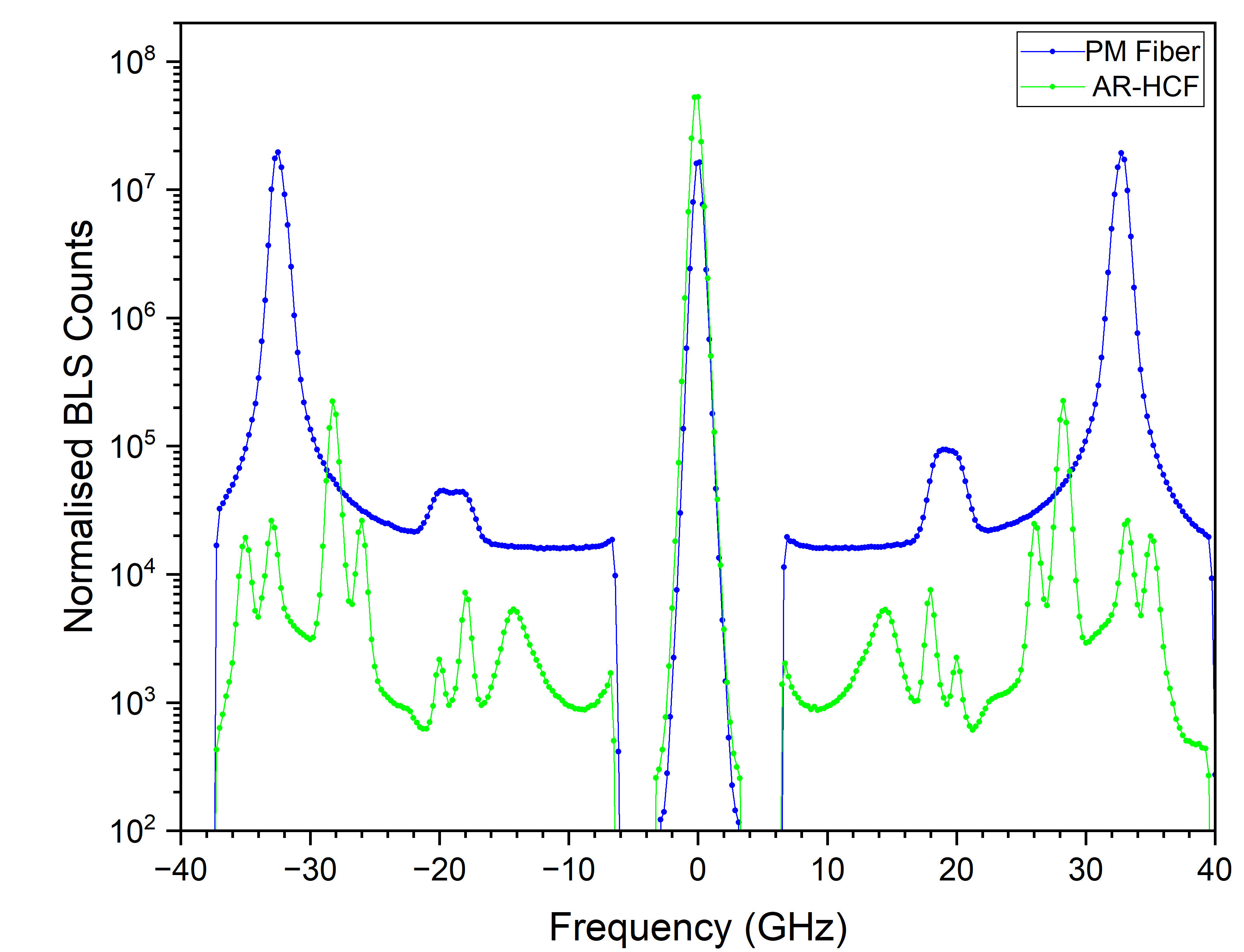}
\caption{\label{Fig3} Spontaneous Brillouin spectra of the PM fiber (Thorlabs PM-S405-XP) (blue) and the AR-HCF (green). The large peak at zero frequency shift corresponds to reference laser line used for calibration and stabilization for the TFPI. Accumulated counts were normalized to transmitted power, fiber length, and number of TFPI scans. The average dark photon count of the detector accumulated during each experiment was subtracted from the corresponding data before normalization.}
\end{figure}

The frequency distribution of the photon count from the inelastically scattered light registered after the TFPI by a single-photon detector is shown in Fig.\,\ref{Fig3}. In the case of the solid-core PM fiber (blue curve), it is possible to identify three peaks which correspond to specific phononic modes, namely longitudinal acoustic (LA $32.6\,\mathrm{GHz}$) and two transverse acoustic (TA, $19.8\,\mathrm{GHz}$ and $18.3\,\mathrm{GHz}$) modes. The appearance of two TA modes is most likely associated with the birefringence induced in the fiber by strain to achieve polarization-maintaining properties. The pronounced feature of the collected spectrum is the broadband optical noise, spanning over the whole range of measured frequencies. The most plausible explanation of the origins of this noise is a glassy and strained structure of the fiber, resulting in a broad manifold of indistinguishable phononic modes. Remarkably, the backscattered signal from the LA mode in the measurement exceeded the intensity of the reference laser line with effective power $<1\,\mathrm{\upmu W}$ used for dynamic frequency stabilization and finesse optimization of the TFPI (see Fig.\,\ref{Fig1}). 

% FIGURE 3

The Brillouin spectrum collected from the AR-HCF is shown in Fig.\,\ref{Fig3} as a green curve. The spectrum contains substantially more peaks than the solid core fiber. However, except for just one peak at $27.81\,\mathrm{GHz}$, the intensity of spectral features is consistently below that of the solid core fiber (up to 3 orders of magnitude, for example at $32.6\,\mathrm{GHz}$), making AR-HCF a good choice for low-noise BLS experiments requiring fiber. However, the large number of different peaks and no particular correspondence of any of those to the bulk phonon spectrum of silica glass raises the question about their nature. To address that, we performed a series of additional measurements as well as numerical simulations.

% FIGURE 4
\begin{figure}
\includegraphics[width=1\columnwidth]{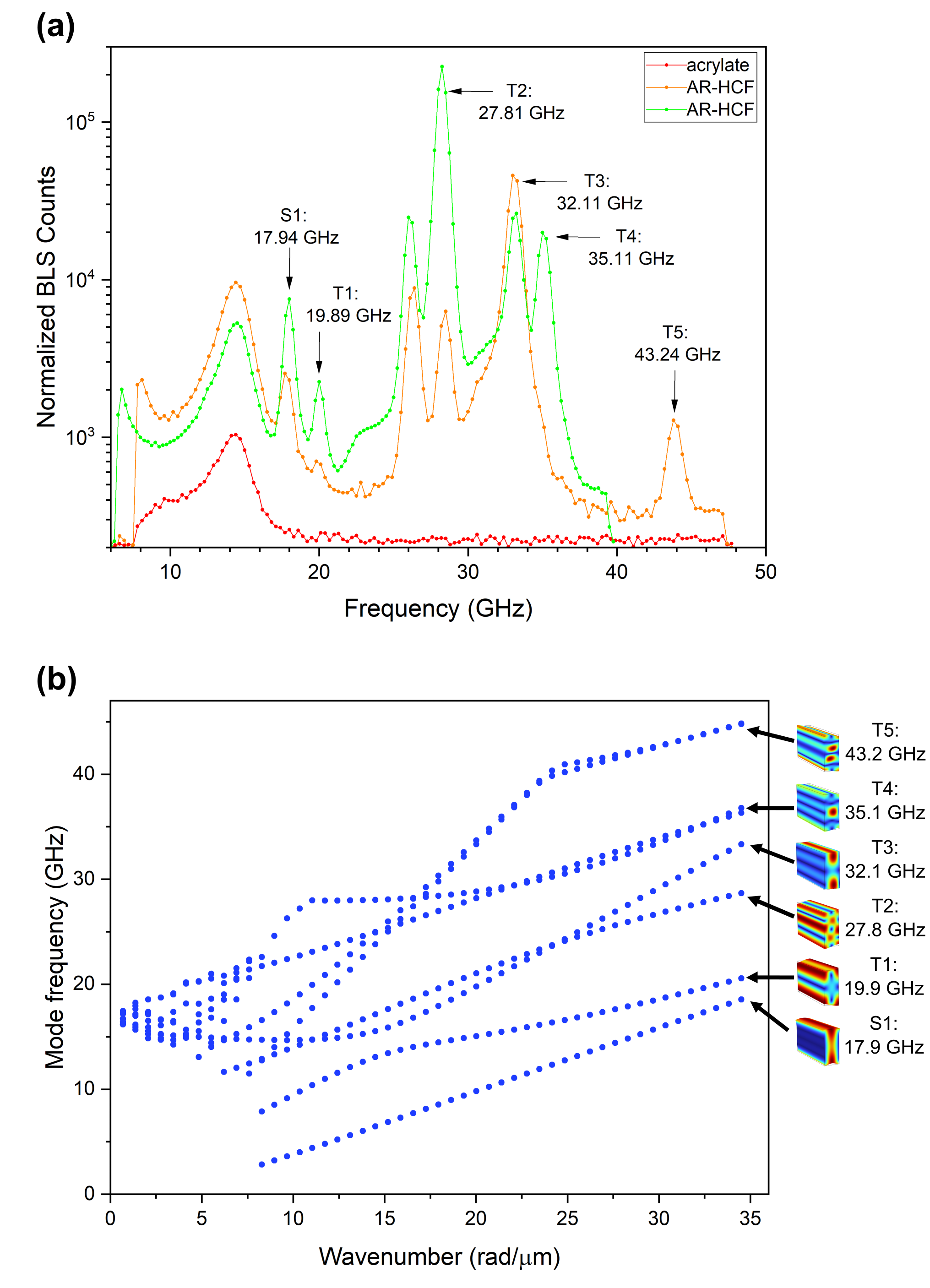}  
\caption{\label{Fig4}(a) Normalized Brillouin anti-Stokes spectrum of the AR-HCF recorded for slightly different alignment (orange and green traces), causing some of the modes to disappear and/or change their intensity. The red trace corresponds to a spectrum collected from a piece of a pure acrylate coating separated from the fiber. The characteristic peak at about $14\,\mathrm{GHz}$ corresponds to bulk LA phonons in acrylate. (b) Numerically simulated dispersion of acoustic modes in a rectangular silica slab of $200\,\mathrm{nm}$ thickness performed using COMSOL Multiphysics. The largest wavenumber in the simulation corresponds to twice the wavevector of the probing light in the backscattering BLS geometry.}
\end{figure}

In Fig.\,\ref{Fig4}a, the BLS anti-Stokes part of the spectrum from the AR-HCF is shown for two different alignments and compared with that from a sample of bulk acrylate taken from the outer coating of the fiber. Comparing those spectra, we can conclude with certainty that below $17\,\mathrm{GHz}$ the only contributor to the BLS signal is the coating with a broad peak centered at about $14\,\mathrm{GHz}$. Most likely, a small portion of the probing light was entering the bulk glass section of the fiber and thus was able to interact with the coating in the first few centimeters of the straight fiber section at the alignment stage. After the first bend in the fiber, all that light would leak out of the fiber and would not be able to contribute to the backscattered light. This contribution could be possibly eliminated by improving the coupling geometry in the following way: One can use an aspheric lens as well as a pinhole directly in front of the fiber end; this can block the light from entering the fiber outside of its guided mode aperture (shown in Fig. \ref{Fig2}).

At higher frequencies, we identify up to seven sharper peaks. In our setup, variations in the probing beam alignment have caused differences in coupling to several of these modes, making the relative peak intensities vary. As evidenced by Fig.\,\ref{Fig4}a, the frequencies at which modes occur are repeatable for different alignments, since they are determined by the boundary conditions and elastic properties of the material itself. These frequencies do not match exactly the dispersion for bulk TA and LA modes in fused silica, but it can be reasonably argued that they originate from thickness modes within the thin capillary structures surrounding the hollow core (see the AR-HCF cross-section shown in Fig.\,\ref{Fig2}), as their dimensions are comparable to the probed acoustic wavelength along the fiber of $\lambda = 182\,\mathrm{nm}$. To reveal the role of confinement on the dispersion relations for the observed acoustic modes in the fused silica substructure, we performed a finite-element simulation of the system using COMSOL Multiphysics. The eigenfrequency study was performed using a simplified geometry of a quasi-infinite slab of $200\,\mathrm{nm}$ thickness with periodic boundary conditions over the other two dimensions. Considering the given scale of the system, it is a reasonable approximation for the capillary geometry. The bulk values of elastic modulus and Poisson ratio parameters for the Heraeus F300 low-OH synthetic silica were set to $72\,\mathrm{GPa}$ and $0.16$ respectively\cite{HeraeusWeb}. 

A sweep across wavevectors between $0$ and $34.51\,\mathrm{rad/\upmu m}$ (the latter corresponds to a probed wavenumber of $4\pi n / \lambda$ for $532\,\mathrm{nm}$ wavelength of the probing optical photons) reveal a dispersion relation for thickness modes shown in Fig.\,\ref{Fig4}b. The simulated slab thickness of $200\,\mathrm{nm}$ results in the eigenfrequencies at the probed wavenumber to match experimentally measured peak frequencies in Fig.\,\ref{Fig4}a. It should be noted that while the simulated thickness value is supported by the direct cross-sectional SEM data (see Fig.\,\ref{Fig2}), there is a difference between these measurements and the thickness estimate of $170-180\,\mathrm{nm}$ from optical measurement of the high-loss resonance bands \cite{Murphy2022}. This discrepancy between optical and SEM measurements is a commonly occurring issue \cite{Jackson2024}, but in our case, it is negligible and does not change the main findings of this study. Considering the confined TA modes' dispersions as well as their mode profiles across the thickness, we can assign more meaningful labels to them. For example, the lowest frequency mode in Fig.\,\ref{Fig4}b (labeled S1) matches close in frequency with a simulated eigenmode with a phase velocity of around $3280\,\mathrm{m/s}$. This mode is in close correspondence with that of a surface wave in bulk fused silica. Such surface waves in solid-core silica fibers are known to have phase velocities between 0.87 and 0.95 of those seen in bulk TA waves.\cite{Beugnot2014} The other peaks seen in the experiment above this mode are labeled T1-T5 correspondingly and also match closely with the simulation. They correspond to the high-order thickness TA modes of capillary walls. The only mode that does not have a corresponding match to the simulation is the unlabeled peak at $26\,\mathrm{GHz}$. Its origins might stem from the other dimensions of confinement, which go outside the scope of this work, such as curvature of the capillary structure, etc. Remarkably, according to the results of the simulation, all the modes observed in the experiment are the TA modes. The possible reason for LA modes' absence is a substantially higher frequency due to confinement pushing them to over $60\,\mathrm{GHz}$ - way over the present measurements frequency range. It is possible to speculate that the traces of a bulk LA mode are visible in the spectrum around $32\,\mathrm{GHz}$, but it is barely resolvable.  \looseness=-1

\section{\label{sec:Conclusion}Conclusions and Outlook}

In this study, we have demonstrated that anti-resonant hollow-core fibers present a significant advantage over conventional solid-core fused silica fibers for applications in spontaneous Brillouin light scattering spectroscopy. By confining light within a predominantly air-filled core, AR-HCFs effectively suppress unwanted phonon-induced inelastic scattering from the fiber material itself. Our experimental results show a reduction in spurious phononic peak intensities by more than two orders of magnitude compared to solid-core fibers, marking a substantial improvement in signal-to-noise ratio---a critical parameter in high-resolution spectroscopic studies.

Despite the presence of some residual spectral features in AR-HCFs, we have identified their likely origins in the structural elements of the fiber, particularly the thin silica capillaries forming the light-guiding core. Finite-element modeling confirms that these features correspond to confined transverse acoustic thickness modes. At the same time, longitudinal acoustic modes appear to be suppressed or shifted beyond the measured frequency range due to geometrical confinement. The ability to identify and interpret confined acoustic modes offers new opportunities to study the elastic properties of microstructured fibers themselves, as well as further improve their design to achieve even better signal-to-noise ratios.  \looseness=-2
\parskip=3pt
These findings establish AR-HCFs as highly suitable candidates for low-noise photonic applications. In particular, integration of AR-HCFs into light scattering experimental setups could open new avenues in fiber-based sensing, precision metrology, quantum communications, and the study of quasiparticle dynamics under extreme conditions, particularly in environments where free-space optics are impractical, such as cryogenic or integrated setups. Furthermore, tailoring the capillary wall geometry and material composition may enable engineered phonon responses, offering a path toward customizable acoustic backgrounds for light-matter interaction studies. 

\begin{acknowledgments}
Authors would like to thank Robert E. Camley and Rair Mac\^{e}do for the fruitful discussions. Research primarily supported by the U.S. Department of Energy (DOE), Office of Science, Basic Energy Sciences (BES) under Award DE-SC0024400 (BLS and SEM measurements, COMSOL simulations). Y.L. acknowledges support by the U.S. Department of Energy, Office of Science, Basic Energy Sciences, Materials Sciences and Engineering Division under Contract No. DE-SC0022060. This work was partially funded by the EPSRC under grant EP/T020903/1 (fiber fabrication).
\end{acknowledgments}

\section*{Data Availability Statement}

The data that support the findings of this study are available within the article and from the corresponding author upon reasonable request.

\nocite{*}
\bibliography{main}% Produces the bibliography via BibTeX.

\providecommand{\noopsort}[1]{}\providecommand{\singleletter}[1]{#1}%
\begin{thebibliography}{70}%
\makeatletter
\providecommand \@ifxundefined [1]{%
 \@ifx{#1\undefined}
}%
\providecommand \@ifnum [1]{%
 \ifnum #1\expandafter \@firstoftwo
 \else \expandafter \@secondoftwo
 \fi
}%
\providecommand \@ifx [1]{%
 \ifx #1\expandafter \@firstoftwo
 \else \expandafter \@secondoftwo
 \fi
}%
\providecommand \natexlab [1]{#1}%
\providecommand \enquote  [1]{``#1''}%
\providecommand \bibnamefont  [1]{#1}%
\providecommand \bibfnamefont [1]{#1}%
\providecommand \citenamefont [1]{#1}%
\providecommand \href@noop [0]{\@secondoftwo}%
\providecommand \href [0]{\begingroup \@sanitize@url \@href}%
\providecommand \@href[1]{\@@startlink{#1}\@@href}%
\providecommand \@@href[1]{\endgroup#1\@@endlink}%
\providecommand \@sanitize@url [0]{\catcode `\\12\catcode `\$12\catcode `\&12\catcode `\#12\catcode `\^12\catcode `\_12\catcode `\%12\relax}%
\providecommand \@@startlink[1]{}%
\providecommand \@@endlink[0]{}%
\providecommand \url  [0]{\begingroup\@sanitize@url \@url }%
\providecommand \@url [1]{\endgroup\@href {#1}{\urlprefix }}%
\providecommand \urlprefix  [0]{URL }%
\providecommand \Eprint [0]{\href }%
\providecommand \doibase [0]{http://dx.doi.org/}%
\providecommand \selectlanguage [0]{\@gobble}%
\providecommand \bibinfo  [0]{\@secondoftwo}%
\providecommand \bibfield  [0]{\@secondoftwo}%
\providecommand \translation [1]{[#1]}%
\providecommand \BibitemOpen [0]{}%
\providecommand \bibitemStop [0]{}%
\providecommand \bibitemNoStop [0]{.\EOS\space}%
\providecommand \EOS [0]{\spacefactor3000\relax}%
\providecommand \BibitemShut  [1]{\csname bibitem#1\endcsname}%
\let\auto@bib@innerbib\@empty
%</preamble>
\bibitem [{\citenamefont {Bozhko}\ \emph {et~al.}(2020)\citenamefont {Bozhko}, \citenamefont {Musiienko-Shmarova}, \citenamefont {Tiberkevich}, \citenamefont {Slavin}, \citenamefont {Syvorotka}, \citenamefont {Hillebrands},\ and\ \citenamefont {Serga}}]{Bozhko2020}%
  \BibitemOpen
  \bibfield  {author} {\bibinfo {author} {\bibfnamefont {D.~A.}\ \bibnamefont {Bozhko}}, \bibinfo {author} {\bibfnamefont {H.~Y.}\ \bibnamefont {Musiienko-Shmarova}}, \bibinfo {author} {\bibfnamefont {V.~S.}\ \bibnamefont {Tiberkevich}}, \bibinfo {author} {\bibfnamefont {A.~N.}\ \bibnamefont {Slavin}}, \bibinfo {author} {\bibfnamefont {I.~I.}\ \bibnamefont {Syvorotka}}, \bibinfo {author} {\bibfnamefont {B.}~\bibnamefont {Hillebrands}}, \ and\ \bibinfo {author} {\bibfnamefont {A.~A.}\ \bibnamefont {Serga}},\ }\bibfield  {title} {\enquote {\bibinfo {title} {Unconventional spin currents in magnetic films},}\ }\href {\doibase 10.1103/physrevresearch.2.023324} {\bibfield  {journal} {\bibinfo  {journal} {Phys. Rev. Res.}\ }\textbf {\bibinfo {volume} {2}},\ \bibinfo {pages} {023324} (\bibinfo {year} {2020})}\BibitemShut {NoStop}%
\bibitem [{\citenamefont {Han}, \citenamefont {Kim},\ and\ \citenamefont {Lee}(2004)}]{Han2004}%
  \BibitemOpen
  \bibfield  {author} {\bibinfo {author} {\bibfnamefont {K.~H.}\ \bibnamefont {Han}}, \bibinfo {author} {\bibfnamefont {J.~G.}\ \bibnamefont {Kim}}, \ and\ \bibinfo {author} {\bibfnamefont {S.}~\bibnamefont {Lee}},\ }\bibfield  {title} {\enquote {\bibinfo {title} {Brillouin light scattering study of the magnetic hysteresis loop in {Fe--Ni/Si(100)} film with induced magnetic anisotropy},}\ }\href {\doibase 10.1016/j.ssc.2003.10.005} {\bibfield  {journal} {\bibinfo  {journal} {Solid State Commun.}\ }\textbf {\bibinfo {volume} {129}},\ \bibinfo {pages} {261--265} (\bibinfo {year} {2004})}\BibitemShut {NoStop}%
\bibitem [{\citenamefont {Sebastian}\ \emph {et~al.}(2015{\natexlab{a}})\citenamefont {Sebastian}, \citenamefont {Kawada}, \citenamefont {Obry}, \citenamefont {Br\"{a}cher}, \citenamefont {Pirro}, \citenamefont {Bozhko}, \citenamefont {Serga}, \citenamefont {Naganuma}, \citenamefont {Oogane}, \citenamefont {Ando},\ and\ \citenamefont {Hillebrands}}]{Sebastian2015}%
  \BibitemOpen
  \bibfield  {author} {\bibinfo {author} {\bibfnamefont {T.}~\bibnamefont {Sebastian}}, \bibinfo {author} {\bibfnamefont {Y.}~\bibnamefont {Kawada}}, \bibinfo {author} {\bibfnamefont {B.}~\bibnamefont {Obry}}, \bibinfo {author} {\bibfnamefont {T.}~\bibnamefont {Br\"{a}cher}}, \bibinfo {author} {\bibfnamefont {P.}~\bibnamefont {Pirro}}, \bibinfo {author} {\bibfnamefont {D.~A.}\ \bibnamefont {Bozhko}}, \bibinfo {author} {\bibfnamefont {A.~A.}\ \bibnamefont {Serga}}, \bibinfo {author} {\bibfnamefont {H.}~\bibnamefont {Naganuma}}, \bibinfo {author} {\bibfnamefont {M.}~\bibnamefont {Oogane}}, \bibinfo {author} {\bibfnamefont {Y.}~\bibnamefont {Ando}}, \ and\ \bibinfo {author} {\bibfnamefont {B.}~\bibnamefont {Hillebrands}},\ }\bibfield  {title} {\enquote {\bibinfo {title} {All-optical characterisation of the spintronic {Heusler} compound $\mathrm{Co}_2\mathrm{Mn}_{0.6}\mathrm{Fe}_{0.4}\mathrm{Si}$},}\ }\href {\doibase 10.1088/0022-3727/48/16/164015} {\bibfield  {journal} {\bibinfo  {journal} {J. Phys. D: Appl.
  Phys.}\ }\textbf {\bibinfo {volume} {48}},\ \bibinfo {pages} {164015} (\bibinfo {year} {2015}{\natexlab{a}})}\BibitemShut {NoStop}%
\bibitem [{\citenamefont {Vavassori}\ \emph {et~al.}(2000)\citenamefont {Vavassori}, \citenamefont {Grimsditch}, \citenamefont {Fullerton}, \citenamefont {Giovannini}, \citenamefont {Zivieri},\ and\ \citenamefont {Nizzoli}}]{Vavassori2000}%
  \BibitemOpen
  \bibfield  {author} {\bibinfo {author} {\bibfnamefont {P.}~\bibnamefont {Vavassori}}, \bibinfo {author} {\bibfnamefont {M.}~\bibnamefont {Grimsditch}}, \bibinfo {author} {\bibfnamefont {E.}~\bibnamefont {Fullerton}}, \bibinfo {author} {\bibfnamefont {L.}~\bibnamefont {Giovannini}}, \bibinfo {author} {\bibfnamefont {R.}~\bibnamefont {Zivieri}}, \ and\ \bibinfo {author} {\bibfnamefont {F.}~\bibnamefont {Nizzoli}},\ }\bibfield  {title} {\enquote {\bibinfo {title} {Brillouin light scattering study of an exchange coupled asymmetric trilayer of {Fe/Cr}},}\ }\href {\doibase 10.1016/s0039-6028(00)00252-1} {\bibfield  {journal} {\bibinfo  {journal} {Surface Science}\ }\textbf {\bibinfo {volume} {454--456}},\ \bibinfo {pages} {880--884} (\bibinfo {year} {2000})}\BibitemShut {NoStop}%
\bibitem [{\citenamefont {Booth}\ \emph {et~al.}(1987)\citenamefont {Booth}, \citenamefont {Srinivasan}, \citenamefont {Patton},\ and\ \citenamefont {de~Gasperis}}]{Booth1987}%
  \BibitemOpen
  \bibfield  {author} {\bibinfo {author} {\bibfnamefont {J.}~\bibnamefont {Booth}}, \bibinfo {author} {\bibfnamefont {G.}~\bibnamefont {Srinivasan}}, \bibinfo {author} {\bibfnamefont {C.}~\bibnamefont {Patton}}, \ and\ \bibinfo {author} {\bibfnamefont {P.}~\bibnamefont {de~Gasperis}},\ }\bibfield  {title} {\enquote {\bibinfo {title} {Brillouin light scattering study of the spin wave stiffness parameter in {Sc}-substituted lutetium-yttrium iron garnet},}\ }\href {\doibase 10.1109/tmag.1987.1065535} {\bibfield  {journal} {\bibinfo  {journal} {IEEE Trans. Magn.}\ }\textbf {\bibinfo {volume} {23}},\ \bibinfo {pages} {3494--3496} (\bibinfo {year} {1987})}\BibitemShut {NoStop}%
\bibitem [{\citenamefont {Zhang}, \citenamefont {Lockwood},\ and\ \citenamefont {Labbe}(1993)}]{Zhang1993}%
  \BibitemOpen
  \bibfield  {author} {\bibinfo {author} {\bibfnamefont {P.~X.}\ \bibnamefont {Zhang}}, \bibinfo {author} {\bibfnamefont {D.~J.}\ \bibnamefont {Lockwood}}, \ and\ \bibinfo {author} {\bibfnamefont {H.~J.}\ \bibnamefont {Labbe}},\ }\bibfield  {title} {\enquote {\bibinfo {title} {Magnon and acoustic-phonon light scattering from {Bi}-doped yttrium iron garnet},}\ }\href {\doibase 10.1103/physrevb.48.6099} {\bibfield  {journal} {\bibinfo  {journal} {Phys. Rev. B}\ }\textbf {\bibinfo {volume} {48}},\ \bibinfo {pages} {6099--6103} (\bibinfo {year} {1993})}\BibitemShut {NoStop}%
\bibitem [{\citenamefont {Belmeguenai}\ \emph {et~al.}(2017)\citenamefont {Belmeguenai}, \citenamefont {Gabor}, \citenamefont {Roussigne}, \citenamefont {Cherif}, \citenamefont {Stashkevich}, \citenamefont {Petrisor}, \citenamefont {Mos},\ and\ \citenamefont {Tiusan}}]{Belmeguenai2017}%
  \BibitemOpen
  \bibfield  {author} {\bibinfo {author} {\bibfnamefont {M.}~\bibnamefont {Belmeguenai}}, \bibinfo {author} {\bibfnamefont {M.~S.}\ \bibnamefont {Gabor}}, \bibinfo {author} {\bibfnamefont {Y.}~\bibnamefont {Roussigne}}, \bibinfo {author} {\bibfnamefont {S.~M.}\ \bibnamefont {Cherif}}, \bibinfo {author} {\bibfnamefont {A.}~\bibnamefont {Stashkevich}}, \bibinfo {author} {\bibfnamefont {T.}~\bibnamefont {Petrisor}}, \bibinfo {author} {\bibfnamefont {R.~B.}\ \bibnamefont {Mos}}, \ and\ \bibinfo {author} {\bibfnamefont {C.}~\bibnamefont {Tiusan}},\ }\bibfield  {title} {\enquote {\bibinfo {title} {{Characterization of the Interfacial Dzyaloshinskii–Moriya Interaction in $\mathrm{Pt/Co_2FeAl_{0.5}Si_{0.5}}$ Ultrathin Films by Brillouin Light Scattering}},}\ }\href {\doibase 10.1109/tmag.2017.2696241} {\bibfield  {journal} {\bibinfo  {journal} {IEEE Transactions on Magnetics}\ }\textbf {\bibinfo {volume} {53}},\ \bibinfo {pages} {1–5} (\bibinfo {year} {2017})}\BibitemShut {NoStop}%
\bibitem [{\citenamefont {Böttcher}\ \emph {et~al.}(2021)\citenamefont {Böttcher}, \citenamefont {Lee}, \citenamefont {Heussner}, \citenamefont {Jaiswal}, \citenamefont {Jakob}, \citenamefont {Kläui}, \citenamefont {Hillebrands}, \citenamefont {Brächer},\ and\ \citenamefont {Pirro}}]{Bottcher2021}%
  \BibitemOpen
  \bibfield  {author} {\bibinfo {author} {\bibfnamefont {T.}~\bibnamefont {Böttcher}}, \bibinfo {author} {\bibfnamefont {K.}~\bibnamefont {Lee}}, \bibinfo {author} {\bibfnamefont {F.}~\bibnamefont {Heussner}}, \bibinfo {author} {\bibfnamefont {S.}~\bibnamefont {Jaiswal}}, \bibinfo {author} {\bibfnamefont {G.}~\bibnamefont {Jakob}}, \bibinfo {author} {\bibfnamefont {M.}~\bibnamefont {Kläui}}, \bibinfo {author} {\bibfnamefont {B.}~\bibnamefont {Hillebrands}}, \bibinfo {author} {\bibfnamefont {T.}~\bibnamefont {Brächer}}, \ and\ \bibinfo {author} {\bibfnamefont {P.}~\bibnamefont {Pirro}},\ }\bibfield  {title} {\enquote {\bibinfo {title} {{Heisenberg Exchange and Dzyaloshinskii–Moriya Interaction in Ultrathin Pt(W)/CoFeB Single and Multilayers}},}\ }\href {\doibase 10.1109/TMAG.2021.3079259} {\bibfield  {journal} {\bibinfo  {journal} {IEEE Transactions on Magnetics}\ }\textbf {\bibinfo {volume} {57}},\ \bibinfo {pages} {1--7} (\bibinfo {year} {2021})}\BibitemShut {NoStop}%
\bibitem [{\citenamefont {Schneider}\ \emph {et~al.}(2013)\citenamefont {Schneider}, \citenamefont {Liaqat}, \citenamefont {El~Boudouti}, \citenamefont {El~Abouti}, \citenamefont {Tremel}, \citenamefont {Butt}, \citenamefont {Djafari-Rouhani},\ and\ \citenamefont {Fytas}}]{Schneider2013}%
  \BibitemOpen
  \bibfield  {author} {\bibinfo {author} {\bibfnamefont {D.}~\bibnamefont {Schneider}}, \bibinfo {author} {\bibfnamefont {F.}~\bibnamefont {Liaqat}}, \bibinfo {author} {\bibfnamefont {E.~H.}\ \bibnamefont {El~Boudouti}}, \bibinfo {author} {\bibfnamefont {O.}~\bibnamefont {El~Abouti}}, \bibinfo {author} {\bibfnamefont {W.}~\bibnamefont {Tremel}}, \bibinfo {author} {\bibfnamefont {H.-J.}\ \bibnamefont {Butt}}, \bibinfo {author} {\bibfnamefont {B.}~\bibnamefont {Djafari-Rouhani}}, \ and\ \bibinfo {author} {\bibfnamefont {G.}~\bibnamefont {Fytas}},\ }\bibfield  {title} {\enquote {\bibinfo {title} {Defect-controlled hypersound propagation in hybrid superlattices},}\ }\href {\doibase 10.1103/physrevlett.111.164301} {\bibfield  {journal} {\bibinfo  {journal} {Phys. Rev. Lett.}\ }\textbf {\bibinfo {volume} {111}},\ \bibinfo {pages} {164301} (\bibinfo {year} {2013})}\BibitemShut {NoStop}%
\bibitem [{\citenamefont {Yudistira}\ \emph {et~al.}(2016)\citenamefont {Yudistira}, \citenamefont {Boes}, \citenamefont {Graczykowski}, \citenamefont {Alzina}, \citenamefont {Yeo}, \citenamefont {Sotomayor~Torres},\ and\ \citenamefont {Mitchell}}]{Yudistira2016}%
  \BibitemOpen
  \bibfield  {author} {\bibinfo {author} {\bibfnamefont {D.}~\bibnamefont {Yudistira}}, \bibinfo {author} {\bibfnamefont {A.}~\bibnamefont {Boes}}, \bibinfo {author} {\bibfnamefont {B.}~\bibnamefont {Graczykowski}}, \bibinfo {author} {\bibfnamefont {F.}~\bibnamefont {Alzina}}, \bibinfo {author} {\bibfnamefont {L.~Y.}\ \bibnamefont {Yeo}}, \bibinfo {author} {\bibfnamefont {C.~M.}\ \bibnamefont {Sotomayor~Torres}}, \ and\ \bibinfo {author} {\bibfnamefont {A.}~\bibnamefont {Mitchell}},\ }\bibfield  {title} {\enquote {\bibinfo {title} {Nanoscale pillar hypersonic surface phononic crystals},}\ }\href {\doibase 10.1103/physrevb.94.094304} {\bibfield  {journal} {\bibinfo  {journal} {Phys. Rev. B}\ }\textbf {\bibinfo {volume} {94}},\ \bibinfo {pages} {094304} (\bibinfo {year} {2016})}\BibitemShut {NoStop}%
\bibitem [{\citenamefont {Ye}\ \emph {et~al.}(2025)\citenamefont {Ye}, \citenamefont {Feng}, \citenamefont {te~Morsche}, \citenamefont {Wei}, \citenamefont {Klaver}, \citenamefont {Mishra}, \citenamefont {Zheng}, \citenamefont {Keloth}, \citenamefont {Tarık~Işık}, \citenamefont {Chen}, \citenamefont {Wang},\ and\ \citenamefont {Marpaung}}]{Ye2025}%
  \BibitemOpen
  \bibfield  {author} {\bibinfo {author} {\bibfnamefont {K.}~\bibnamefont {Ye}}, \bibinfo {author} {\bibfnamefont {H.}~\bibnamefont {Feng}}, \bibinfo {author} {\bibfnamefont {R.}~\bibnamefont {te~Morsche}}, \bibinfo {author} {\bibfnamefont {C.}~\bibnamefont {Wei}}, \bibinfo {author} {\bibfnamefont {Y.}~\bibnamefont {Klaver}}, \bibinfo {author} {\bibfnamefont {A.}~\bibnamefont {Mishra}}, \bibinfo {author} {\bibfnamefont {Z.}~\bibnamefont {Zheng}}, \bibinfo {author} {\bibfnamefont {A.}~\bibnamefont {Keloth}}, \bibinfo {author} {\bibfnamefont {A.}~\bibnamefont {Tarık~Işık}}, \bibinfo {author} {\bibfnamefont {Z.}~\bibnamefont {Chen}}, \bibinfo {author} {\bibfnamefont {C.}~\bibnamefont {Wang}}, \ and\ \bibinfo {author} {\bibfnamefont {D.}~\bibnamefont {Marpaung}},\ }\bibfield  {title} {\enquote {\bibinfo {title} {Integrated brillouin photonics in thin-film lithium niobate},}\ }\href {\doibase 10.1126/sciadv.adv4022} {\bibfield  {journal} {\bibinfo  {journal} {Science Advances}\ }\textbf {\bibinfo {volume} {11}}
  (\bibinfo {year} {2025}),\ 10.1126/sciadv.adv4022}\BibitemShut {NoStop}%
\bibitem [{\citenamefont {Fohr}\ \emph {et~al.}(2009)\citenamefont {Fohr}, \citenamefont {Serga}, \citenamefont {Schneider}, \citenamefont {Hamrle},\ and\ \citenamefont {Hillebrands}}]{Fohr2009}%
  \BibitemOpen
  \bibfield  {author} {\bibinfo {author} {\bibfnamefont {F.}~\bibnamefont {Fohr}}, \bibinfo {author} {\bibfnamefont {A.~A.}\ \bibnamefont {Serga}}, \bibinfo {author} {\bibfnamefont {T.}~\bibnamefont {Schneider}}, \bibinfo {author} {\bibfnamefont {J.}~\bibnamefont {Hamrle}}, \ and\ \bibinfo {author} {\bibfnamefont {B.}~\bibnamefont {Hillebrands}},\ }\bibfield  {title} {\enquote {\bibinfo {title} {{Phase sensitive Brillouin scattering measurements with a novel magneto-optic modulator}},}\ }\href {\doibase 10.1063/1.3115210} {\bibfield  {journal} {\bibinfo  {journal} {Rev. Sci. Instrum.}\ }\textbf {\bibinfo {volume} {80}},\ \bibinfo {pages} {043903} (\bibinfo {year} {2009})}\BibitemShut {NoStop}%
\bibitem [{\citenamefont {Vogt}\ \emph {et~al.}(2009)\citenamefont {Vogt}, \citenamefont {Schultheiss}, \citenamefont {Hermsdoerfer}, \citenamefont {Pirro}, \citenamefont {Serga},\ and\ \citenamefont {Hillebrands}}]{Vogt2009}%
  \BibitemOpen
  \bibfield  {author} {\bibinfo {author} {\bibfnamefont {K.}~\bibnamefont {Vogt}}, \bibinfo {author} {\bibfnamefont {H.}~\bibnamefont {Schultheiss}}, \bibinfo {author} {\bibfnamefont {S.~J.}\ \bibnamefont {Hermsdoerfer}}, \bibinfo {author} {\bibfnamefont {P.}~\bibnamefont {Pirro}}, \bibinfo {author} {\bibfnamefont {A.~A.}\ \bibnamefont {Serga}}, \ and\ \bibinfo {author} {\bibfnamefont {B.}~\bibnamefont {Hillebrands}},\ }\bibfield  {title} {\enquote {\bibinfo {title} {{All-optical detection of phase fronts of propagating spin waves in a Ni$_{81}$Fe$_{19}$ microstripe}},}\ }\href {\doibase 10.1063/1.3262348} {\bibfield  {journal} {\bibinfo  {journal} {Appl. Phys. Lett.}\ }\textbf {\bibinfo {volume} {95}},\ \bibinfo {pages} {182508} (\bibinfo {year} {2009})}\BibitemShut {NoStop}%
\bibitem [{\citenamefont {Demidov}\ \emph {et~al.}(2004)\citenamefont {Demidov}, \citenamefont {Demokritov}, \citenamefont {Hillebrands}, \citenamefont {Laufenberg},\ and\ \citenamefont {Freitas}}]{Demidov2004}%
  \BibitemOpen
  \bibfield  {author} {\bibinfo {author} {\bibfnamefont {V.~E.}\ \bibnamefont {Demidov}}, \bibinfo {author} {\bibfnamefont {S.~O.}\ \bibnamefont {Demokritov}}, \bibinfo {author} {\bibfnamefont {B.}~\bibnamefont {Hillebrands}}, \bibinfo {author} {\bibfnamefont {M.}~\bibnamefont {Laufenberg}}, \ and\ \bibinfo {author} {\bibfnamefont {P.~P.}\ \bibnamefont {Freitas}},\ }\bibfield  {title} {\enquote {\bibinfo {title} {Radiation of spin waves by a single micrometer-sized magnetic element},}\ }\href {\doibase 10.1063/1.1803621} {\bibfield  {journal} {\bibinfo  {journal} {Appl. Phys. Lett.}\ }\textbf {\bibinfo {volume} {85}},\ \bibinfo {pages} {2866--2868} (\bibinfo {year} {2004})}\BibitemShut {NoStop}%
\bibitem [{\citenamefont {Demidov}\ and\ \citenamefont {Demokritov}(2015)}]{Demidov2015}%
  \BibitemOpen
  \bibfield  {author} {\bibinfo {author} {\bibfnamefont {V.~E.}\ \bibnamefont {Demidov}}\ and\ \bibinfo {author} {\bibfnamefont {S.~O.}\ \bibnamefont {Demokritov}},\ }\bibfield  {title} {\enquote {\bibinfo {title} {Magnonic waveguides studied by microfocus {Brillouin} light scattering},}\ }\href {\doibase 10.1109/tmag.2014.2388196} {\bibfield  {journal} {\bibinfo  {journal} {IEEE Trans. Magn.}\ }\textbf {\bibinfo {volume} {51}},\ \bibinfo {pages} {1--15} (\bibinfo {year} {2015})}\BibitemShut {NoStop}%
\bibitem [{\citenamefont {Madami}\ \emph {et~al.}(2011)\citenamefont {Madami}, \citenamefont {Bonetti}, \citenamefont {Consolo}, \citenamefont {Tacchi}, \citenamefont {Carlotti}, \citenamefont {Gubbiotti}, \citenamefont {Mancoff}, \citenamefont {Yar},\ and\ \citenamefont {Åkerman}}]{Madami2011}%
  \BibitemOpen
  \bibfield  {author} {\bibinfo {author} {\bibfnamefont {M.}~\bibnamefont {Madami}}, \bibinfo {author} {\bibfnamefont {S.}~\bibnamefont {Bonetti}}, \bibinfo {author} {\bibfnamefont {G.}~\bibnamefont {Consolo}}, \bibinfo {author} {\bibfnamefont {S.}~\bibnamefont {Tacchi}}, \bibinfo {author} {\bibfnamefont {G.}~\bibnamefont {Carlotti}}, \bibinfo {author} {\bibfnamefont {G.}~\bibnamefont {Gubbiotti}}, \bibinfo {author} {\bibfnamefont {F.~B.}\ \bibnamefont {Mancoff}}, \bibinfo {author} {\bibfnamefont {M.~A.}\ \bibnamefont {Yar}}, \ and\ \bibinfo {author} {\bibfnamefont {J.}~\bibnamefont {Åkerman}},\ }\bibfield  {title} {\enquote {\bibinfo {title} {Direct observation of a propagating spin wave induced by spin-transfer torque},}\ }\href {\doibase 10.1038/nnano.2011.140} {\bibfield  {journal} {\bibinfo  {journal} {Nat. Nanotechnol.}\ }\textbf {\bibinfo {volume} {6}},\ \bibinfo {pages} {635--638} (\bibinfo {year} {2011})}\BibitemShut {NoStop}%
\bibitem [{\citenamefont {Heussner}\ \emph {et~al.}(2020)\citenamefont {Heussner}, \citenamefont {Talmelli}, \citenamefont {Geilen}, \citenamefont {Heinz}, \citenamefont {Br\"{a}cher}, \citenamefont {Meyer}, \citenamefont {Ciubotaru}, \citenamefont {Adelmann}, \citenamefont {Yamamoto}, \citenamefont {Serga}, \citenamefont {Hillebrands},\ and\ \citenamefont {Pirro}}]{Heussner2020}%
  \BibitemOpen
  \bibfield  {author} {\bibinfo {author} {\bibfnamefont {F.}~\bibnamefont {Heussner}}, \bibinfo {author} {\bibfnamefont {G.}~\bibnamefont {Talmelli}}, \bibinfo {author} {\bibfnamefont {M.}~\bibnamefont {Geilen}}, \bibinfo {author} {\bibfnamefont {B.}~\bibnamefont {Heinz}}, \bibinfo {author} {\bibfnamefont {T.}~\bibnamefont {Br\"{a}cher}}, \bibinfo {author} {\bibfnamefont {T.}~\bibnamefont {Meyer}}, \bibinfo {author} {\bibfnamefont {F.}~\bibnamefont {Ciubotaru}}, \bibinfo {author} {\bibfnamefont {C.}~\bibnamefont {Adelmann}}, \bibinfo {author} {\bibfnamefont {K.}~\bibnamefont {Yamamoto}}, \bibinfo {author} {\bibfnamefont {A.~A.}\ \bibnamefont {Serga}}, \bibinfo {author} {\bibfnamefont {B.}~\bibnamefont {Hillebrands}}, \ and\ \bibinfo {author} {\bibfnamefont {P.}~\bibnamefont {Pirro}},\ }\bibfield  {title} {\enquote {\bibinfo {title} {Experimental realization of a passive gigahertz frequency‐division demultiplexer for magnonic logic networks},}\ }\href {\doibase 10.1002/pssr.201900695} {\bibfield  {journal}
  {\bibinfo  {journal} {Phys. Status Solidi RRL}\ }\textbf {\bibinfo {volume} {14}},\ \bibinfo {pages} {1900695} (\bibinfo {year} {2020})}\BibitemShut {NoStop}%
\bibitem [{\citenamefont {Dunagin}, \citenamefont {Serga},\ and\ \citenamefont {Bozhko}(2025)}]{Dunagin2025}%
  \BibitemOpen
  \bibfield  {author} {\bibinfo {author} {\bibfnamefont {R.~E.}\ \bibnamefont {Dunagin}}, \bibinfo {author} {\bibfnamefont {A.~A.}\ \bibnamefont {Serga}}, \ and\ \bibinfo {author} {\bibfnamefont {D.~A.}\ \bibnamefont {Bozhko}},\ }\bibfield  {title} {\enquote {\bibinfo {title} {Brillouin light scattering spectroscopy of magnon–phonon thermal spectra of an in-plane magnetized {YIG} film in two-dimensional wavevector space},}\ }\href {\doibase 10.1063/5.0251149} {\bibfield  {journal} {\bibinfo  {journal} {Journal of Applied Physics}\ }\textbf {\bibinfo {volume} {137}} (\bibinfo {year} {2025}),\ 10.1063/5.0251149}\BibitemShut {NoStop}%
\bibitem [{\citenamefont {Hillebrands}, \citenamefont {Baumgart},\ and\ \citenamefont {G\"{u}ntherodt}(1989)}]{Hillebrands1989}%
  \BibitemOpen
  \bibfield  {author} {\bibinfo {author} {\bibfnamefont {B.}~\bibnamefont {Hillebrands}}, \bibinfo {author} {\bibfnamefont {P.}~\bibnamefont {Baumgart}}, \ and\ \bibinfo {author} {\bibfnamefont {G.}~\bibnamefont {G\"{u}ntherodt}},\ }\bibfield  {title} {\enquote {\bibinfo {title} {Brillouin light scattering from spin waves in magnetic layers and multilayers},}\ }\href {\doibase 10.1007/bf00616984} {\bibfield  {journal} {\bibinfo  {journal} {Appl. Phys. A}\ }\textbf {\bibinfo {volume} {49}},\ \bibinfo {pages} {589--598} (\bibinfo {year} {1989})}\BibitemShut {NoStop}%
\bibitem [{\citenamefont {B\"{u}ttner}\ \emph {et~al.}(2000)\citenamefont {B\"{u}ttner}, \citenamefont {Bauer}, \citenamefont {Demokritov}, \citenamefont {Hillebrands}, \citenamefont {Kivshar}, \citenamefont {Grimalsky}, \citenamefont {Rapoport},\ and\ \citenamefont {Slavin}}]{Buettner2000}%
  \BibitemOpen
  \bibfield  {author} {\bibinfo {author} {\bibfnamefont {O.}~\bibnamefont {B\"{u}ttner}}, \bibinfo {author} {\bibfnamefont {M.}~\bibnamefont {Bauer}}, \bibinfo {author} {\bibfnamefont {S.~O.}\ \bibnamefont {Demokritov}}, \bibinfo {author} {\bibfnamefont {B.}~\bibnamefont {Hillebrands}}, \bibinfo {author} {\bibfnamefont {Y.~S.}\ \bibnamefont {Kivshar}}, \bibinfo {author} {\bibfnamefont {V.}~\bibnamefont {Grimalsky}}, \bibinfo {author} {\bibfnamefont {Y.}~\bibnamefont {Rapoport}}, \ and\ \bibinfo {author} {\bibfnamefont {A.~N.}\ \bibnamefont {Slavin}},\ }\bibfield  {title} {\enquote {\bibinfo {title} {Linear and nonlinear diffraction of dipolar spin waves in yttrium iron garnet films observed by space- and time-resolved {Brillouin} light scattering},}\ }\href {\doibase 10.1103/physrevb.61.11576} {\bibfield  {journal} {\bibinfo  {journal} {Phys. Rev. B}\ }\textbf {\bibinfo {volume} {61}},\ \bibinfo {pages} {11576--11587} (\bibinfo {year} {2000})}\BibitemShut {NoStop}%
\bibitem [{\citenamefont {Sebastian}\ \emph {et~al.}(2015{\natexlab{b}})\citenamefont {Sebastian}, \citenamefont {Schultheiss}, \citenamefont {Obry}, \citenamefont {Hillebrands},\ and\ \citenamefont {Schultheiss}}]{Sebastian2015microBLS}%
  \BibitemOpen
  \bibfield  {author} {\bibinfo {author} {\bibfnamefont {T.}~\bibnamefont {Sebastian}}, \bibinfo {author} {\bibfnamefont {K.}~\bibnamefont {Schultheiss}}, \bibinfo {author} {\bibfnamefont {B.}~\bibnamefont {Obry}}, \bibinfo {author} {\bibfnamefont {B.}~\bibnamefont {Hillebrands}}, \ and\ \bibinfo {author} {\bibfnamefont {H.}~\bibnamefont {Schultheiss}},\ }\bibfield  {title} {\enquote {\bibinfo {title} {Micro-focused {Brillouin} light scattering: imaging spin waves at the nanoscale},}\ }\href {\doibase 10.3389/fphy.2015.00035} {\bibfield  {journal} {\bibinfo  {journal} {Front. Phys.}\ }\textbf {\bibinfo {volume} {3}},\ \bibinfo {pages} {35} (\bibinfo {year} {2015}{\natexlab{b}})}\BibitemShut {NoStop}%
\bibitem [{\citenamefont {Xia}\ \emph {et~al.}(1998)\citenamefont {Xia}, \citenamefont {Kabos}, \citenamefont {Zhang}, \citenamefont {Kolodin},\ and\ \citenamefont {Patton}}]{Xia1998}%
  \BibitemOpen
  \bibfield  {author} {\bibinfo {author} {\bibfnamefont {H.}~\bibnamefont {Xia}}, \bibinfo {author} {\bibfnamefont {P.}~\bibnamefont {Kabos}}, \bibinfo {author} {\bibfnamefont {H.~Y.}\ \bibnamefont {Zhang}}, \bibinfo {author} {\bibfnamefont {P.~A.}\ \bibnamefont {Kolodin}}, \ and\ \bibinfo {author} {\bibfnamefont {C.~E.}\ \bibnamefont {Patton}},\ }\bibfield  {title} {\enquote {\bibinfo {title} {Brillouin light scattering and magnon wave vector distributions for microwave-magnetic-envelope solitons in yttrium-iron-garnet thin films},}\ }\href {\doibase 10.1103/physrevlett.81.449} {\bibfield  {journal} {\bibinfo  {journal} {Phys. Rev. Lett.}\ }\textbf {\bibinfo {volume} {81}},\ \bibinfo {pages} {449--452} (\bibinfo {year} {1998})}\BibitemShut {NoStop}%
\bibitem [{\citenamefont {Hahn}\ \emph {et~al.}(2022)\citenamefont {Hahn}, \citenamefont {Frey}, \citenamefont {Serga}, \citenamefont {Vasyuchka}, \citenamefont {Hillebrands}, \citenamefont {Kopietz},\ and\ \citenamefont {R\"{u}ckriegel}}]{Hahn2022}%
  \BibitemOpen
  \bibfield  {author} {\bibinfo {author} {\bibfnamefont {V.}~\bibnamefont {Hahn}}, \bibinfo {author} {\bibfnamefont {P.}~\bibnamefont {Frey}}, \bibinfo {author} {\bibfnamefont {A.~A.}\ \bibnamefont {Serga}}, \bibinfo {author} {\bibfnamefont {V.~I.}\ \bibnamefont {Vasyuchka}}, \bibinfo {author} {\bibfnamefont {B.}~\bibnamefont {Hillebrands}}, \bibinfo {author} {\bibfnamefont {P.}~\bibnamefont {Kopietz}}, \ and\ \bibinfo {author} {\bibfnamefont {A.}~\bibnamefont {R\"{u}ckriegel}},\ }\bibfield  {title} {\enquote {\bibinfo {title} {Accumulation of magnetoelastic bosons in yttrium iron garnet: {Kinetic} theory and wave-vector-resolved {Brillouin} light scattering},}\ }\href {\doibase 10.1103/physrevb.105.144421} {\bibfield  {journal} {\bibinfo  {journal} {Phys. Rev. B}\ }\textbf {\bibinfo {volume} {105}},\ \bibinfo {pages} {144421} (\bibinfo {year} {2022})}\BibitemShut {NoStop}%
\bibitem [{\citenamefont {Kunz}\ \emph {et~al.}(2024)\citenamefont {Kunz}, \citenamefont {K\"{u}ß}, \citenamefont {Schneider}, \citenamefont {Geilen}, \citenamefont {Pirro}, \citenamefont {Albrecht},\ and\ \citenamefont {Weiler}}]{Kunz2024}%
  \BibitemOpen
  \bibfield  {author} {\bibinfo {author} {\bibfnamefont {Y.}~\bibnamefont {Kunz}}, \bibinfo {author} {\bibfnamefont {M.}~\bibnamefont {K\"{u}ß}}, \bibinfo {author} {\bibfnamefont {M.}~\bibnamefont {Schneider}}, \bibinfo {author} {\bibfnamefont {M.}~\bibnamefont {Geilen}}, \bibinfo {author} {\bibfnamefont {P.}~\bibnamefont {Pirro}}, \bibinfo {author} {\bibfnamefont {M.}~\bibnamefont {Albrecht}}, \ and\ \bibinfo {author} {\bibfnamefont {M.}~\bibnamefont {Weiler}},\ }\bibfield  {title} {\enquote {\bibinfo {title} {Coherent surface acoustic wave--spin wave interactions detected by micro-focused {Brillouin} light scattering spectroscopy},}\ }\href {\doibase 10.1063/5.0189324} {\bibfield  {journal} {\bibinfo  {journal} {Appl. Phys. Lett.}\ }\textbf {\bibinfo {volume} {124}},\ \bibinfo {pages} {152403} (\bibinfo {year} {2024})}\BibitemShut {NoStop}%
\bibitem [{\citenamefont {Frey}\ \emph {et~al.}(2021)\citenamefont {Frey}, \citenamefont {Bozhko}, \citenamefont {L’vov}, \citenamefont {Hillebrands},\ and\ \citenamefont {Serga}}]{Frey2021}%
  \BibitemOpen
  \bibfield  {author} {\bibinfo {author} {\bibfnamefont {P.}~\bibnamefont {Frey}}, \bibinfo {author} {\bibfnamefont {D.~A.}\ \bibnamefont {Bozhko}}, \bibinfo {author} {\bibfnamefont {V.~S.}\ \bibnamefont {L’vov}}, \bibinfo {author} {\bibfnamefont {B.}~\bibnamefont {Hillebrands}}, \ and\ \bibinfo {author} {\bibfnamefont {A.~A.}\ \bibnamefont {Serga}},\ }\bibfield  {title} {\enquote {\bibinfo {title} {Double accumulation and anisotropic transport of magnetoelastic bosons in yttrium iron garnet films},}\ }\href {\doibase 10.1103/physrevb.104.014420} {\bibfield  {journal} {\bibinfo  {journal} {Phys. Rev. B}\ }\textbf {\bibinfo {volume} {104}},\ \bibinfo {pages} {014420} (\bibinfo {year} {2021})}\BibitemShut {NoStop}%
\bibitem [{\citenamefont {Ord\'{o}\~{n}ez Romero}\ \emph {et~al.}(2009)\citenamefont {Ord\'{o}\~{n}ez Romero}, \citenamefont {Kalinikos}, \citenamefont {Krivosik}, \citenamefont {Tong}, \citenamefont {Kabos},\ and\ \citenamefont {Patton}}]{OrdezRomero2009}%
  \BibitemOpen
  \bibfield  {author} {\bibinfo {author} {\bibfnamefont {C.~L.}\ \bibnamefont {Ord\'{o}\~{n}ez Romero}}, \bibinfo {author} {\bibfnamefont {B.~A.}\ \bibnamefont {Kalinikos}}, \bibinfo {author} {\bibfnamefont {P.}~\bibnamefont {Krivosik}}, \bibinfo {author} {\bibfnamefont {W.}~\bibnamefont {Tong}}, \bibinfo {author} {\bibfnamefont {P.}~\bibnamefont {Kabos}}, \ and\ \bibinfo {author} {\bibfnamefont {C.~E.}\ \bibnamefont {Patton}},\ }\bibfield  {title} {\enquote {\bibinfo {title} {Three-magnon splitting and confluence processes for spin-wave excitations in yttrium iron garnet films: {Wave} vector selective {Brillouin} light scattering measurements and analysis},}\ }\href {\doibase 10.1103/physrevb.79.144428} {\bibfield  {journal} {\bibinfo  {journal} {Phys. Rev. B}\ }\textbf {\bibinfo {volume} {79}},\ \bibinfo {pages} {144428} (\bibinfo {year} {2009})}\BibitemShut {NoStop}%
\bibitem [{\citenamefont {Sandercock}\ and\ \citenamefont {Wettling}(1973)}]{Sandercock1973}%
  \BibitemOpen
  \bibfield  {author} {\bibinfo {author} {\bibfnamefont {J.}~\bibnamefont {Sandercock}}\ and\ \bibinfo {author} {\bibfnamefont {W.}~\bibnamefont {Wettling}},\ }\bibfield  {title} {\enquote {\bibinfo {title} {{Light scattering from thermal acoustic magnons in yttrium iron garnet}},}\ }\href {\doibase 10.1016/0038-1098(73)90276-7} {\bibfield  {journal} {\bibinfo  {journal} {Solid State Commun.}\ }\textbf {\bibinfo {volume} {13}},\ \bibinfo {pages} {1729--1732} (\bibinfo {year} {1973})}\BibitemShut {NoStop}%
\bibitem [{\citenamefont {Buchmeier}\ \emph {et~al.}(2007)\citenamefont {Buchmeier}, \citenamefont {Dassow}, \citenamefont {B\"{u}rgler},\ and\ \citenamefont {Schneider}}]{Buchmeier2007}%
  \BibitemOpen
  \bibfield  {author} {\bibinfo {author} {\bibfnamefont {M.}~\bibnamefont {Buchmeier}}, \bibinfo {author} {\bibfnamefont {H.}~\bibnamefont {Dassow}}, \bibinfo {author} {\bibfnamefont {D.~E.}\ \bibnamefont {B\"{u}rgler}}, \ and\ \bibinfo {author} {\bibfnamefont {C.~M.}\ \bibnamefont {Schneider}},\ }\bibfield  {title} {\enquote {\bibinfo {title} {Intensity of {Brillouin} light scattering from spin waves in magnetic multilayers with noncollinear spin configurations: {Theory} and experiment},}\ }\href {\doibase 10.1103/physrevb.75.184436} {\bibfield  {journal} {\bibinfo  {journal} {Physical Review B}\ }\textbf {\bibinfo {volume} {75}} (\bibinfo {year} {2007}),\ 10.1103/physrevb.75.184436}\BibitemShut {NoStop}%
\bibitem [{\citenamefont {Wettling}, \citenamefont {Cottam},\ and\ \citenamefont {Sandercock}(1975)}]{Wettling1975}%
  \BibitemOpen
  \bibfield  {author} {\bibinfo {author} {\bibfnamefont {W.}~\bibnamefont {Wettling}}, \bibinfo {author} {\bibfnamefont {M.~G.}\ \bibnamefont {Cottam}}, \ and\ \bibinfo {author} {\bibfnamefont {J.~R.}\ \bibnamefont {Sandercock}},\ }\bibfield  {title} {\enquote {\bibinfo {title} {The relation between one-magnon light scattering and the complex magneto-optic effects in {YIG}},}\ }\href {\doibase 10.1088/0022-3719/8/2/014} {\bibfield  {journal} {\bibinfo  {journal} {J. Phys. C: Solid State Phys.}\ }\textbf {\bibinfo {volume} {8}},\ \bibinfo {pages} {211--228} (\bibinfo {year} {1975})}\BibitemShut {NoStop}%
\bibitem [{\citenamefont {Cryer-Jenkins}\ \emph {et~al.}(2025)\citenamefont {Cryer-Jenkins}, \citenamefont {Leung}, \citenamefont {Rathee}, \citenamefont {Tan}, \citenamefont {Major},\ and\ \citenamefont {Vanner}}]{CryerJenkins2025}%
  \BibitemOpen
  \bibfield  {author} {\bibinfo {author} {\bibfnamefont {E.~A.}\ \bibnamefont {Cryer-Jenkins}}, \bibinfo {author} {\bibfnamefont {A.~C.}\ \bibnamefont {Leung}}, \bibinfo {author} {\bibfnamefont {H.}~\bibnamefont {Rathee}}, \bibinfo {author} {\bibfnamefont {A.~K.~C.}\ \bibnamefont {Tan}}, \bibinfo {author} {\bibfnamefont {K.~D.}\ \bibnamefont {Major}}, \ and\ \bibinfo {author} {\bibfnamefont {M.~R.}\ \bibnamefont {Vanner}},\ }\bibfield  {title} {\enquote {\bibinfo {title} {Brillouin–mandelstam scattering in telecommunications optical fiber at millikelvin temperatures},}\ }\href {\doibase 10.1063/5.0241253} {\bibfield  {journal} {\bibinfo  {journal} {APL Photonics}\ }\textbf {\bibinfo {volume} {10}} (\bibinfo {year} {2025}),\ 10.1063/5.0241253}\BibitemShut {NoStop}%
\bibitem [{\citenamefont {Mizuno}\ and\ \citenamefont {Nakamura}(2010)}]{Mizuno2010}%
  \BibitemOpen
  \bibfield  {author} {\bibinfo {author} {\bibfnamefont {Y.}~\bibnamefont {Mizuno}}\ and\ \bibinfo {author} {\bibfnamefont {K.}~\bibnamefont {Nakamura}},\ }\bibfield  {title} {\enquote {\bibinfo {title} {Experimental study of brillouin scattering in perfluorinated polymer optical fiber at telecommunication wavelength},}\ }\href {\doibase 10.1063/1.3463038} {\bibfield  {journal} {\bibinfo  {journal} {Applied Physics Letters}\ }\textbf {\bibinfo {volume} {97}} (\bibinfo {year} {2010}),\ 10.1063/1.3463038}\BibitemShut {NoStop}%
\bibitem [{\citenamefont {Kosareva}\ \emph {et~al.}(2022)\citenamefont {Kosareva}, \citenamefont {Alyukova}, \citenamefont {Salnikov},\ and\ \citenamefont {Kalinin}}]{Kosareva2022}%
  \BibitemOpen
  \bibfield  {author} {\bibinfo {author} {\bibfnamefont {A.}~\bibnamefont {Kosareva}}, \bibinfo {author} {\bibfnamefont {V.}~\bibnamefont {Alyukova}}, \bibinfo {author} {\bibfnamefont {N.}~\bibnamefont {Salnikov}}, \ and\ \bibinfo {author} {\bibfnamefont {N.}~\bibnamefont {Kalinin}},\ }\bibfield  {title} {\enquote {\bibinfo {title} {Numerical study of stimulated brillouin scattering in optical microcavities made of telecommunication fibres},}\ }in\ \href {\doibase 10.1109/nusod54938.2022.9894744} {\emph {\bibinfo {booktitle} {2022 International Conference on Numerical Simulation of Optoelectronic Devices (NUSOD)}}}\ (\bibinfo  {publisher} {IEEE},\ \bibinfo {year} {2022})\ p.\ \bibinfo {pages} {183–184}\BibitemShut {NoStop}%
\bibitem [{\citenamefont {Ji}\ \emph {et~al.}(2024)\citenamefont {Ji}, \citenamefont {Huang}, \citenamefont {He}, \citenamefont {Yin}, \citenamefont {Zheng}, \citenamefont {Jiang}, \citenamefont {Leng},\ and\ \citenamefont {Pang}}]{Ji2024}%
  \BibitemOpen
  \bibfield  {author} {\bibinfo {author} {\bibfnamefont {G.}~\bibnamefont {Ji}}, \bibinfo {author} {\bibfnamefont {Z.}~\bibnamefont {Huang}}, \bibinfo {author} {\bibfnamefont {W.}~\bibnamefont {He}}, \bibinfo {author} {\bibfnamefont {R.}~\bibnamefont {Yin}}, \bibinfo {author} {\bibfnamefont {Y.}~\bibnamefont {Zheng}}, \bibinfo {author} {\bibfnamefont {X.}~\bibnamefont {Jiang}}, \bibinfo {author} {\bibfnamefont {Y.}~\bibnamefont {Leng}}, \ and\ \bibinfo {author} {\bibfnamefont {M.}~\bibnamefont {Pang}},\ }\bibfield  {title} {\enquote {\bibinfo {title} {Kilohertz-linewidth brillouin photonic crystal fiber laser with non-resonance pumping configuration},}\ }\href {\doibase 10.1109/jphot.2024.3399030} {\bibfield  {journal} {\bibinfo  {journal} {IEEE Photonics Journal}\ }\textbf {\bibinfo {volume} {16}},\ \bibinfo {pages} {1–6} (\bibinfo {year} {2024})}\BibitemShut {NoStop}%
\bibitem [{\citenamefont {Chauhan}\ \emph {et~al.}(2021)\citenamefont {Chauhan}, \citenamefont {Isichenko}, \citenamefont {Liu}, \citenamefont {Wang}, \citenamefont {Zhao}, \citenamefont {Behunin}, \citenamefont {Rakich}, \citenamefont {Jayich}, \citenamefont {Fertig}, \citenamefont {Hoyt},\ and\ \citenamefont {Blumenthal}}]{Chauhan2021}%
  \BibitemOpen
  \bibfield  {author} {\bibinfo {author} {\bibfnamefont {N.}~\bibnamefont {Chauhan}}, \bibinfo {author} {\bibfnamefont {A.}~\bibnamefont {Isichenko}}, \bibinfo {author} {\bibfnamefont {K.}~\bibnamefont {Liu}}, \bibinfo {author} {\bibfnamefont {J.}~\bibnamefont {Wang}}, \bibinfo {author} {\bibfnamefont {Q.}~\bibnamefont {Zhao}}, \bibinfo {author} {\bibfnamefont {R.~O.}\ \bibnamefont {Behunin}}, \bibinfo {author} {\bibfnamefont {P.~T.}\ \bibnamefont {Rakich}}, \bibinfo {author} {\bibfnamefont {A.~M.}\ \bibnamefont {Jayich}}, \bibinfo {author} {\bibfnamefont {C.}~\bibnamefont {Fertig}}, \bibinfo {author} {\bibfnamefont {C.~W.}\ \bibnamefont {Hoyt}}, \ and\ \bibinfo {author} {\bibfnamefont {D.~J.}\ \bibnamefont {Blumenthal}},\ }\bibfield  {title} {\enquote {\bibinfo {title} {{Visible light photonic integrated Brillouin laser}},}\ }\href {\doibase 10.1038/s41467-021-24926-8} {\bibfield  {journal} {\bibinfo  {journal} {Nature Communications}\ }\textbf {\bibinfo {volume} {12}} (\bibinfo {year} {2021}),\
  10.1038/s41467-021-24926-8}\BibitemShut {NoStop}%
\bibitem [{\citenamefont {Liu}\ \emph {et~al.}(2021)\citenamefont {Liu}, \citenamefont {Choudhary}, \citenamefont {Ren}, \citenamefont {Choi}, \citenamefont {Casas‐Bedoya}, \citenamefont {Morrison}, \citenamefont {Ma}, \citenamefont {Nguyen}, \citenamefont {Mitchell}, \citenamefont {Madden}, \citenamefont {Marpaung},\ and\ \citenamefont {Eggleton}}]{Liu2021}%
  \BibitemOpen
  \bibfield  {author} {\bibinfo {author} {\bibfnamefont {Y.}~\bibnamefont {Liu}}, \bibinfo {author} {\bibfnamefont {A.}~\bibnamefont {Choudhary}}, \bibinfo {author} {\bibfnamefont {G.}~\bibnamefont {Ren}}, \bibinfo {author} {\bibfnamefont {D.}~\bibnamefont {Choi}}, \bibinfo {author} {\bibfnamefont {A.}~\bibnamefont {Casas‐Bedoya}}, \bibinfo {author} {\bibfnamefont {B.}~\bibnamefont {Morrison}}, \bibinfo {author} {\bibfnamefont {P.}~\bibnamefont {Ma}}, \bibinfo {author} {\bibfnamefont {T.~G.}\ \bibnamefont {Nguyen}}, \bibinfo {author} {\bibfnamefont {A.}~\bibnamefont {Mitchell}}, \bibinfo {author} {\bibfnamefont {S.~J.}\ \bibnamefont {Madden}}, \bibinfo {author} {\bibfnamefont {D.}~\bibnamefont {Marpaung}}, \ and\ \bibinfo {author} {\bibfnamefont {B.~J.}\ \bibnamefont {Eggleton}},\ }\bibfield  {title} {\enquote {\bibinfo {title} {Circulator‐free {Brillouin} photonic planar circuit},}\ }\href {\doibase 10.1002/lpor.202000481} {\bibfield  {journal} {\bibinfo  {journal} {Laser and Photonics Reviews}\ }\textbf
  {\bibinfo {volume} {15}} (\bibinfo {year} {2021}),\ 10.1002/lpor.202000481}\BibitemShut {NoStop}%
\bibitem [{\citenamefont {Gundavarapu}\ \emph {et~al.}(2018)\citenamefont {Gundavarapu}, \citenamefont {Brodnik}, \citenamefont {Puckett}, \citenamefont {Huffman}, \citenamefont {Bose}, \citenamefont {Behunin}, \citenamefont {Wu}, \citenamefont {Qiu}, \citenamefont {Pinho}, \citenamefont {Chauhan}, \citenamefont {Nohava}, \citenamefont {Rakich}, \citenamefont {Nelson}, \citenamefont {Salit},\ and\ \citenamefont {Blumenthal}}]{Gundavarapu2018}%
  \BibitemOpen
  \bibfield  {author} {\bibinfo {author} {\bibfnamefont {S.}~\bibnamefont {Gundavarapu}}, \bibinfo {author} {\bibfnamefont {G.~M.}\ \bibnamefont {Brodnik}}, \bibinfo {author} {\bibfnamefont {M.}~\bibnamefont {Puckett}}, \bibinfo {author} {\bibfnamefont {T.}~\bibnamefont {Huffman}}, \bibinfo {author} {\bibfnamefont {D.}~\bibnamefont {Bose}}, \bibinfo {author} {\bibfnamefont {R.}~\bibnamefont {Behunin}}, \bibinfo {author} {\bibfnamefont {J.}~\bibnamefont {Wu}}, \bibinfo {author} {\bibfnamefont {T.}~\bibnamefont {Qiu}}, \bibinfo {author} {\bibfnamefont {C.}~\bibnamefont {Pinho}}, \bibinfo {author} {\bibfnamefont {N.}~\bibnamefont {Chauhan}}, \bibinfo {author} {\bibfnamefont {J.}~\bibnamefont {Nohava}}, \bibinfo {author} {\bibfnamefont {P.~T.}\ \bibnamefont {Rakich}}, \bibinfo {author} {\bibfnamefont {K.~D.}\ \bibnamefont {Nelson}}, \bibinfo {author} {\bibfnamefont {M.}~\bibnamefont {Salit}}, \ and\ \bibinfo {author} {\bibfnamefont {D.~J.}\ \bibnamefont {Blumenthal}},\ }\bibfield  {title} {\enquote {\bibinfo
  {title} {{Sub-hertz fundamental linewidth photonic integrated Brillouin laser}},}\ }\href {\doibase 10.1038/s41566-018-0313-2} {\bibfield  {journal} {\bibinfo  {journal} {Nature Photonics}\ }\textbf {\bibinfo {volume} {13}},\ \bibinfo {pages} {60–67} (\bibinfo {year} {2018})}\BibitemShut {NoStop}%
\bibitem [{\citenamefont {H\"{o}gele}\ \emph {et~al.}(2008)\citenamefont {H\"{o}gele}, \citenamefont {Seidl}, \citenamefont {Kroner}, \citenamefont {Karrai}, \citenamefont {Schulhauser}, \citenamefont {Sqalli}, \citenamefont {Scrimgeour},\ and\ \citenamefont {Warburton}}]{Hgele2008}%
  \BibitemOpen
  \bibfield  {author} {\bibinfo {author} {\bibfnamefont {A.}~\bibnamefont {H\"{o}gele}}, \bibinfo {author} {\bibfnamefont {S.}~\bibnamefont {Seidl}}, \bibinfo {author} {\bibfnamefont {M.}~\bibnamefont {Kroner}}, \bibinfo {author} {\bibfnamefont {K.}~\bibnamefont {Karrai}}, \bibinfo {author} {\bibfnamefont {C.}~\bibnamefont {Schulhauser}}, \bibinfo {author} {\bibfnamefont {O.}~\bibnamefont {Sqalli}}, \bibinfo {author} {\bibfnamefont {J.}~\bibnamefont {Scrimgeour}}, \ and\ \bibinfo {author} {\bibfnamefont {R.~J.}\ \bibnamefont {Warburton}},\ }\bibfield  {title} {\enquote {\bibinfo {title} {Fiber-based confocal microscope for cryogenic spectroscopy},}\ }\href {\doibase 10.1063/1.2885681} {\bibfield  {journal} {\bibinfo  {journal} {Review of Scientific Instruments}\ }\textbf {\bibinfo {volume} {79}} (\bibinfo {year} {2008}),\ 10.1063/1.2885681}\BibitemShut {NoStop}%
\bibitem [{\citenamefont {Christiansen}\ \emph {et~al.}(2023)\citenamefont {Christiansen}, \citenamefont {Popelka}, \citenamefont {Gom}, \citenamefont {Naylor},\ and\ \citenamefont {Stolov}}]{Christiansen2023}%
  \BibitemOpen
  \bibfield  {author} {\bibinfo {author} {\bibfnamefont {A.~J.}\ \bibnamefont {Christiansen}}, \bibinfo {author} {\bibfnamefont {M.}~\bibnamefont {Popelka}}, \bibinfo {author} {\bibfnamefont {B.~G.}\ \bibnamefont {Gom}}, \bibinfo {author} {\bibfnamefont {D.~A.}\ \bibnamefont {Naylor}}, \ and\ \bibinfo {author} {\bibfnamefont {A.~A.}\ \bibnamefont {Stolov}},\ }\bibfield  {title} {\enquote {\bibinfo {title} {Characterization of optical fiber at cryogenic temperatures},}\ }in\ \href {\doibase 10.1117/12.2647281} {\emph {\bibinfo {booktitle} {Optical Components and Materials XX}}},\ \bibinfo {editor} {edited by\ \bibinfo {editor} {\bibfnamefont {M.~J.}\ \bibnamefont {Digonnet}}\ and\ \bibinfo {editor} {\bibfnamefont {S.}~\bibnamefont {Jiang}}}\ (\bibinfo  {publisher} {SPIE},\ \bibinfo {year} {2023})\ p.~\bibinfo {pages} {35}\BibitemShut {NoStop}%
\bibitem [{\citenamefont {Tokizaki}\ \emph {et~al.}(1999)\citenamefont {Tokizaki}, \citenamefont {Sugiyama}, \citenamefont {Onuki},\ and\ \citenamefont {Tani}}]{Tokizaki1999}%
  \BibitemOpen
  \bibfield  {author} {\bibinfo {author} {\bibfnamefont {T.}~\bibnamefont {Tokizaki}}, \bibinfo {author} {\bibfnamefont {K.}~\bibnamefont {Sugiyama}}, \bibinfo {author} {\bibfnamefont {T.}~\bibnamefont {Onuki}}, \ and\ \bibinfo {author} {\bibfnamefont {T.}~\bibnamefont {Tani}},\ }\bibfield  {title} {\enquote {\bibinfo {title} {Optical‐fibre scanning near‐field optical microscope for cryogenic operation},}\ }\href {\doibase 10.1046/j.1365-2818.1999.00522.x} {\bibfield  {journal} {\bibinfo  {journal} {Journal of Microscopy}\ }\textbf {\bibinfo {volume} {194}},\ \bibinfo {pages} {321–324} (\bibinfo {year} {1999})}\BibitemShut {NoStop}%
\bibitem [{\citenamefont {Mahar}\ \emph {et~al.}(2008)\citenamefont {Mahar}, \citenamefont {Geng}, \citenamefont {Schultz}, \citenamefont {Minervini}, \citenamefont {Jiang}, \citenamefont {Titus}, \citenamefont {Takayasu}, \citenamefont {yu~Gung}, \citenamefont {Tian},\ and\ \citenamefont {Chavez-Pirson}}]{Mahar2008}%
  \BibitemOpen
  \bibfield  {author} {\bibinfo {author} {\bibfnamefont {S.}~\bibnamefont {Mahar}}, \bibinfo {author} {\bibfnamefont {J.}~\bibnamefont {Geng}}, \bibinfo {author} {\bibfnamefont {J.}~\bibnamefont {Schultz}}, \bibinfo {author} {\bibfnamefont {J.}~\bibnamefont {Minervini}}, \bibinfo {author} {\bibfnamefont {S.}~\bibnamefont {Jiang}}, \bibinfo {author} {\bibfnamefont {P.}~\bibnamefont {Titus}}, \bibinfo {author} {\bibfnamefont {M.}~\bibnamefont {Takayasu}}, \bibinfo {author} {\bibfnamefont {C.}~\bibnamefont {yu~Gung}}, \bibinfo {author} {\bibfnamefont {W.}~\bibnamefont {Tian}}, \ and\ \bibinfo {author} {\bibfnamefont {A.}~\bibnamefont {Chavez-Pirson}},\ }\bibfield  {title} {\enquote {\bibinfo {title} {{Real-time simultaneous temperature and strain measurements at cryogenic temperatures in an optical fiber}},}\ }in\ \href {\doibase 10.1117/12.791913} {\emph {\bibinfo {booktitle} {Remote Sensing System Engineering}}},\ Vol.\ \bibinfo {volume} {7087},\ \bibinfo {editor} {edited by\ \bibinfo {editor} {\bibfnamefont
  {P.~E.}\ \bibnamefont {Ardanuy}}\ and\ \bibinfo {editor} {\bibfnamefont {J.~J.}\ \bibnamefont {Puschell}}},\ \bibinfo {organization} {International Society for Optics and Photonics}\ (\bibinfo  {publisher} {SPIE},\ \bibinfo {year} {2008})\ p.\ \bibinfo {pages} {70870I}\BibitemShut {NoStop}%
\bibitem [{\citenamefont {Beugnot}\ \emph {et~al.}(2014)\citenamefont {Beugnot}, \citenamefont {Lebrun}, \citenamefont {Pauliat}, \citenamefont {Maillotte}, \citenamefont {Laude},\ and\ \citenamefont {Sylvestre}}]{Beugnot2014}%
  \BibitemOpen
  \bibfield  {author} {\bibinfo {author} {\bibfnamefont {J.-C.}\ \bibnamefont {Beugnot}}, \bibinfo {author} {\bibfnamefont {S.}~\bibnamefont {Lebrun}}, \bibinfo {author} {\bibfnamefont {G.}~\bibnamefont {Pauliat}}, \bibinfo {author} {\bibfnamefont {H.}~\bibnamefont {Maillotte}}, \bibinfo {author} {\bibfnamefont {V.}~\bibnamefont {Laude}}, \ and\ \bibinfo {author} {\bibfnamefont {T.}~\bibnamefont {Sylvestre}},\ }\bibfield  {title} {\enquote {\bibinfo {title} {Brillouin light scattering from surface acoustic waves in a subwavelength-diameter optical fibre},}\ }\href {\doibase 10.1038/ncomms6242} {\bibfield  {journal} {\bibinfo  {journal} {Nature Communications}\ }\textbf {\bibinfo {volume} {5}} (\bibinfo {year} {2014}),\ 10.1038/ncomms6242}\BibitemShut {NoStop}%
\bibitem [{\citenamefont {Tan}, \citenamefont {Huang},\ and\ \citenamefont {Huang}(2006)}]{Tan2006}%
  \BibitemOpen
  \bibfield  {author} {\bibinfo {author} {\bibfnamefont {K.~J.}\ \bibnamefont {Tan}}, \bibinfo {author} {\bibfnamefont {C.~Z.}\ \bibnamefont {Huang}}, \ and\ \bibinfo {author} {\bibfnamefont {Y.~M.}\ \bibnamefont {Huang}},\ }\bibfield  {title} {\enquote {\bibinfo {title} {Determination of lead in environmental water by a backward light scattering technique},}\ }\href {\doibase https://doi.org/10.1016/j.talanta.2005.12.027} {\bibfield  {journal} {\bibinfo  {journal} {Talanta}\ }\textbf {\bibinfo {volume} {70}},\ \bibinfo {pages} {116--121} (\bibinfo {year} {2006})},\ \bibinfo {note} {china-Japan-Korea Environmental Analytical Chemistry Symposium}\BibitemShut {NoStop}%
\bibitem [{\citenamefont {Shirasaki}\ and\ \citenamefont {Haus}(1992)}]{Shirasaki1992}%
  \BibitemOpen
  \bibfield  {author} {\bibinfo {author} {\bibfnamefont {M.}~\bibnamefont {Shirasaki}}\ and\ \bibinfo {author} {\bibfnamefont {H.~A.}\ \bibnamefont {Haus}},\ }\bibfield  {title} {\enquote {\bibinfo {title} {Reduction of guided-acoustic-wave {Brillouin} scattering noise in a squeezer},}\ }\href {\doibase 10.1364/ol.17.001225} {\bibfield  {journal} {\bibinfo  {journal} {Optics Letters}\ }\textbf {\bibinfo {volume} {17}},\ \bibinfo {pages} {1225} (\bibinfo {year} {1992})}\BibitemShut {NoStop}%
\bibitem [{\citenamefont {Bergman}\ and\ \citenamefont {Haus}(1991)}]{Bergman1991}%
  \BibitemOpen
  \bibfield  {author} {\bibinfo {author} {\bibfnamefont {K.}~\bibnamefont {Bergman}}\ and\ \bibinfo {author} {\bibfnamefont {H.~A.}\ \bibnamefont {Haus}},\ }\bibfield  {title} {\enquote {\bibinfo {title} {Squeezing in fibers with optical pulses},}\ }\href {\doibase 10.1364/ol.16.000663} {\bibfield  {journal} {\bibinfo  {journal} {Optics Letters}\ }\textbf {\bibinfo {volume} {16}},\ \bibinfo {pages} {663} (\bibinfo {year} {1991})}\BibitemShut {NoStop}%
\bibitem [{\citenamefont {Shelby}\ \emph {et~al.}(1986)\citenamefont {Shelby}, \citenamefont {Levenson}, \citenamefont {Perlmutter}, \citenamefont {DeVoe},\ and\ \citenamefont {Walls}}]{Shelby1986}%
  \BibitemOpen
  \bibfield  {author} {\bibinfo {author} {\bibfnamefont {R.~M.}\ \bibnamefont {Shelby}}, \bibinfo {author} {\bibfnamefont {M.~D.}\ \bibnamefont {Levenson}}, \bibinfo {author} {\bibfnamefont {S.~H.}\ \bibnamefont {Perlmutter}}, \bibinfo {author} {\bibfnamefont {R.~G.}\ \bibnamefont {DeVoe}}, \ and\ \bibinfo {author} {\bibfnamefont {D.~F.}\ \bibnamefont {Walls}},\ }\bibfield  {title} {\enquote {\bibinfo {title} {Broad-band parametric deamplification of quantum noise in an optical fiber},}\ }\href {\doibase 10.1103/physrevlett.57.691} {\bibfield  {journal} {\bibinfo  {journal} {Physical Review Letters}\ }\textbf {\bibinfo {volume} {57}},\ \bibinfo {pages} {691–694} (\bibinfo {year} {1986})}\BibitemShut {NoStop}%
\bibitem [{\citenamefont {Bergman}\ \emph {et~al.}(1993)\citenamefont {Bergman}, \citenamefont {Doerr}, \citenamefont {Haus},\ and\ \citenamefont {Shirasaki}}]{Bergman1993}%
  \BibitemOpen
  \bibfield  {author} {\bibinfo {author} {\bibfnamefont {K.}~\bibnamefont {Bergman}}, \bibinfo {author} {\bibfnamefont {C.~R.}\ \bibnamefont {Doerr}}, \bibinfo {author} {\bibfnamefont {H.~A.}\ \bibnamefont {Haus}}, \ and\ \bibinfo {author} {\bibfnamefont {M.}~\bibnamefont {Shirasaki}},\ }\bibfield  {title} {\enquote {\bibinfo {title} {Sub-shot-noise measurement with fiber-squeezed optical pulses},}\ }\href {\doibase 10.1364/ol.18.000643} {\bibfield  {journal} {\bibinfo  {journal} {Optics Letters}\ }\textbf {\bibinfo {volume} {18}},\ \bibinfo {pages} {643} (\bibinfo {year} {1993})}\BibitemShut {NoStop}%
\bibitem [{\citenamefont {Feng}\ \emph {et~al.}(2017)\citenamefont {Feng}, \citenamefont {Wan}, \citenamefont {Li},\ and\ \citenamefont {Zhang}}]{Feng2017}%
  \BibitemOpen
  \bibfield  {author} {\bibinfo {author} {\bibfnamefont {J.}~\bibnamefont {Feng}}, \bibinfo {author} {\bibfnamefont {Z.}~\bibnamefont {Wan}}, \bibinfo {author} {\bibfnamefont {Y.}~\bibnamefont {Li}}, \ and\ \bibinfo {author} {\bibfnamefont {K.}~\bibnamefont {Zhang}},\ }\bibfield  {title} {\enquote {\bibinfo {title} {Distribution of continuous variable quantum entanglement at a telecommunication wavelength over 20km of optical fiber},}\ }\href {\doibase 10.1364/ol.42.003399} {\bibfield  {journal} {\bibinfo  {journal} {Optics Letters}\ }\textbf {\bibinfo {volume} {42}},\ \bibinfo {pages} {3399} (\bibinfo {year} {2017})}\BibitemShut {NoStop}%
\bibitem [{\citenamefont {Lassen}\ \emph {et~al.}(2013)\citenamefont {Lassen}, \citenamefont {Berni}, \citenamefont {Madsen}, \citenamefont {Filip},\ and\ \citenamefont {Andersen}}]{Lassen2013}%
  \BibitemOpen
  \bibfield  {author} {\bibinfo {author} {\bibfnamefont {M.}~\bibnamefont {Lassen}}, \bibinfo {author} {\bibfnamefont {A.}~\bibnamefont {Berni}}, \bibinfo {author} {\bibfnamefont {L.~S.}\ \bibnamefont {Madsen}}, \bibinfo {author} {\bibfnamefont {R.}~\bibnamefont {Filip}}, \ and\ \bibinfo {author} {\bibfnamefont {U.~L.}\ \bibnamefont {Andersen}},\ }\bibfield  {title} {\enquote {\bibinfo {title} {Gaussian error correction of quantum states in a correlated noisy channel},}\ }\href {\doibase 10.1103/physrevlett.111.180502} {\bibfield  {journal} {\bibinfo  {journal} {Physical Review Letters}\ }\textbf {\bibinfo {volume} {111}} (\bibinfo {year} {2013}),\ 10.1103/physrevlett.111.180502}\BibitemShut {NoStop}%
\bibitem [{\citenamefont {Li}\ \emph {et~al.}(2014)\citenamefont {Li}, \citenamefont {Wang}, \citenamefont {Wang},\ and\ \citenamefont {Bai}}]{Li2014}%
  \BibitemOpen
  \bibfield  {author} {\bibinfo {author} {\bibfnamefont {Y.}~\bibnamefont {Li}}, \bibinfo {author} {\bibfnamefont {N.}~\bibnamefont {Wang}}, \bibinfo {author} {\bibfnamefont {X.}~\bibnamefont {Wang}}, \ and\ \bibinfo {author} {\bibfnamefont {Z.}~\bibnamefont {Bai}},\ }\bibfield  {title} {\enquote {\bibinfo {title} {Influence of guided acoustic wave brillouin scattering on excess noise in fiber-based continuous variable quantum key distribution},}\ }\href {\doibase 10.1364/josab.31.002379} {\bibfield  {journal} {\bibinfo  {journal} {Journal of the Optical Society of America B}\ }\textbf {\bibinfo {volume} {31}},\ \bibinfo {pages} {2379} (\bibinfo {year} {2014})}\BibitemShut {NoStop}%
\bibitem [{\citenamefont {Fokoua}\ \emph {et~al.}(2023)\citenamefont {Fokoua}, \citenamefont {Mousavi}, \citenamefont {Jasion}, \citenamefont {Richardson},\ and\ \citenamefont {Poletti}}]{Fokoua2023}%
  \BibitemOpen
  \bibfield  {author} {\bibinfo {author} {\bibfnamefont {E.~N.}\ \bibnamefont {Fokoua}}, \bibinfo {author} {\bibfnamefont {S.~A.}\ \bibnamefont {Mousavi}}, \bibinfo {author} {\bibfnamefont {G.~T.}\ \bibnamefont {Jasion}}, \bibinfo {author} {\bibfnamefont {D.~J.}\ \bibnamefont {Richardson}}, \ and\ \bibinfo {author} {\bibfnamefont {F.}~\bibnamefont {Poletti}},\ }\bibfield  {title} {\enquote {\bibinfo {title} {Loss in hollow-core optical fibers: mechanisms, scaling rules, and limits},}\ }\href {\doibase 10.1364/AOP.470592} {\bibfield  {journal} {\bibinfo  {journal} {Adv. Opt. Photon.}\ }\textbf {\bibinfo {volume} {15}},\ \bibinfo {pages} {1--85} (\bibinfo {year} {2023})}\BibitemShut {NoStop}%
\bibitem [{\citenamefont {Mears}\ \emph {et~al.}(2024)\citenamefont {Mears}, \citenamefont {Harrington}, \citenamefont {Wadsworth}, \citenamefont {Knight}, \citenamefont {Stone},\ and\ \citenamefont {Birks}}]{Mears2024_guidance}%
  \BibitemOpen
  \bibfield  {author} {\bibinfo {author} {\bibfnamefont {R.}~\bibnamefont {Mears}}, \bibinfo {author} {\bibfnamefont {K.}~\bibnamefont {Harrington}}, \bibinfo {author} {\bibfnamefont {W.~J.}\ \bibnamefont {Wadsworth}}, \bibinfo {author} {\bibfnamefont {J.~C.}\ \bibnamefont {Knight}}, \bibinfo {author} {\bibfnamefont {J.~M.}\ \bibnamefont {Stone}}, \ and\ \bibinfo {author} {\bibfnamefont {T.~A.}\ \bibnamefont {Birks}},\ }\bibfield  {title} {\enquote {\bibinfo {title} {Guidance of ultraviolet light down to 190 nm in a hollow-core optical fibre},}\ }\href {\doibase 10.1364/OE.509212} {\bibfield  {journal} {\bibinfo  {journal} {Opt. Express}\ }\textbf {\bibinfo {volume} {32}},\ \bibinfo {pages} {8520--8526} (\bibinfo {year} {2024})}\BibitemShut {NoStop}%
\bibitem [{\citenamefont {Iyer}\ \emph {et~al.}(2020)\citenamefont {Iyer}, \citenamefont {Xu}, \citenamefont {Antonio-Lopez}, \citenamefont {Correa},\ and\ \citenamefont {Renninger}}]{Iyer2020}%
  \BibitemOpen
  \bibfield  {author} {\bibinfo {author} {\bibfnamefont {A.}~\bibnamefont {Iyer}}, \bibinfo {author} {\bibfnamefont {W.}~\bibnamefont {Xu}}, \bibinfo {author} {\bibfnamefont {J.~E.}\ \bibnamefont {Antonio-Lopez}}, \bibinfo {author} {\bibfnamefont {R.~A.}\ \bibnamefont {Correa}}, \ and\ \bibinfo {author} {\bibfnamefont {W.~H.}\ \bibnamefont {Renninger}},\ }\bibfield  {title} {\enquote {\bibinfo {title} {Ultra-low {Brillouin} scattering in anti-resonant hollow-core fibers},}\ }\href {\doibase 10.1063/5.0017796} {\bibfield  {journal} {\bibinfo  {journal} {APL Photonics}\ }\textbf {\bibinfo {volume} {5}},\ \bibinfo {pages} {096109} (\bibinfo {year} {2020})}\BibitemShut {NoStop}%
\bibitem [{\citenamefont {Renninger}\ \emph {et~al.}(2016)\citenamefont {Renninger}, \citenamefont {Shin}, \citenamefont {Behunin}, \citenamefont {Kharel}, \citenamefont {Kittlaus},\ and\ \citenamefont {Rakich}}]{Renninger2016}%
  \BibitemOpen
  \bibfield  {author} {\bibinfo {author} {\bibfnamefont {W.~H.}\ \bibnamefont {Renninger}}, \bibinfo {author} {\bibfnamefont {H.}~\bibnamefont {Shin}}, \bibinfo {author} {\bibfnamefont {R.~O.}\ \bibnamefont {Behunin}}, \bibinfo {author} {\bibfnamefont {P.}~\bibnamefont {Kharel}}, \bibinfo {author} {\bibfnamefont {E.~A.}\ \bibnamefont {Kittlaus}}, \ and\ \bibinfo {author} {\bibfnamefont {P.~T.}\ \bibnamefont {Rakich}},\ }\bibfield  {title} {\enquote {\bibinfo {title} {Forward {Brillouin} scattering in hollow-core photonic bandgap fibers},}\ }\href {\doibase 10.1088/1367-2630/18/2/025008} {\bibfield  {journal} {\bibinfo  {journal} {New Journal of Physics}\ }\textbf {\bibinfo {volume} {18}},\ \bibinfo {pages} {025008} (\bibinfo {year} {2016})}\BibitemShut {NoStop}%
\bibitem [{\citenamefont {Elser}\ \emph {et~al.}(2015)\citenamefont {Elser}, \citenamefont {Stiller}, \citenamefont {Elser}, \citenamefont {Heim}, \citenamefont {Marquardt},\ and\ \citenamefont {Leuchs}}]{Zhong2015}%
  \BibitemOpen
  \bibfield  {author} {\bibinfo {author} {\bibfnamefont {W.}~\bibnamefont {Elser}}, \bibinfo {author} {\bibfnamefont {B.}~\bibnamefont {Stiller}}, \bibinfo {author} {\bibfnamefont {D.}~\bibnamefont {Elser}}, \bibinfo {author} {\bibfnamefont {B.}~\bibnamefont {Heim}}, \bibinfo {author} {\bibfnamefont {C.}~\bibnamefont {Marquardt}}, \ and\ \bibinfo {author} {\bibfnamefont {G.}~\bibnamefont {Leuchs}},\ }\bibfield  {title} {\enquote {\bibinfo {title} {Depolarized guided acoustic wave {Brillouin} scattering in hollow-core photonic crystal fibers},}\ }\href {\doibase 10.1364/oe.23.027707} {\bibfield  {journal} {\bibinfo  {journal} {Optics Express}\ }\textbf {\bibinfo {volume} {23}},\ \bibinfo {pages} {27707} (\bibinfo {year} {2015})}\BibitemShut {NoStop}%
\bibitem [{\citenamefont {Cardona}\ \emph {et~al.}(2000)\citenamefont {Cardona}, \citenamefont {G\"{u}ntherodt}, \citenamefont {Hillebrands},\ and\ \citenamefont {G.}}]{Cardona2000}%
  \BibitemOpen
  \bibfield  {author} {\bibinfo {author} {\bibfnamefont {M.}~\bibnamefont {Cardona}}, \bibinfo {author} {\bibfnamefont {G.}~\bibnamefont {G\"{u}ntherodt}}, \bibinfo {author} {\bibfnamefont {B.}~\bibnamefont {Hillebrands}}, \ and\ \bibinfo {author} {\bibfnamefont {S.}~\bibnamefont {G.}},\ }\href {\doibase 10.1007/bfb0103383} {\emph {\bibinfo {title} {{Light Scattering in Solids VII}}}},\ \bibinfo {number} {204}\ (\bibinfo  {publisher} {Springer Berlin Heidelberg},\ \bibinfo {year} {2000})\BibitemShut {NoStop}%
\bibitem [{\citenamefont {Merklein}\ \emph {et~al.}(2022)\citenamefont {Merklein}, \citenamefont {Kabakova}, \citenamefont {Zarifi},\ and\ \citenamefont {Eggleton}}]{Merklein2022}%
  \BibitemOpen
  \bibfield  {author} {\bibinfo {author} {\bibfnamefont {M.}~\bibnamefont {Merklein}}, \bibinfo {author} {\bibfnamefont {I.~V.}\ \bibnamefont {Kabakova}}, \bibinfo {author} {\bibfnamefont {A.}~\bibnamefont {Zarifi}}, \ and\ \bibinfo {author} {\bibfnamefont {B.~J.}\ \bibnamefont {Eggleton}},\ }\bibfield  {title} {\enquote {\bibinfo {title} {100 years of {Brillouin} scattering: Historical and future perspectives},}\ }\href {\doibase 10.1063/5.0095488} {\bibfield  {journal} {\bibinfo  {journal} {Applied Physics Reviews}\ }\textbf {\bibinfo {volume} {9}} (\bibinfo {year} {2022}),\ 10.1063/5.0095488}\BibitemShut {NoStop}%
\bibitem [{\citenamefont {Chiao}, \citenamefont {Townes},\ and\ \citenamefont {Stoicheff}(1964)}]{Chiao1964}%
  \BibitemOpen
  \bibfield  {author} {\bibinfo {author} {\bibfnamefont {R.~Y.}\ \bibnamefont {Chiao}}, \bibinfo {author} {\bibfnamefont {C.~H.}\ \bibnamefont {Townes}}, \ and\ \bibinfo {author} {\bibfnamefont {B.~P.}\ \bibnamefont {Stoicheff}},\ }\bibfield  {title} {\enquote {\bibinfo {title} {Stimulated {Brillouin} scattering and coherent generation of intense hypersonic waves},}\ }\href {\doibase 10.1103/physrevlett.12.592} {\bibfield  {journal} {\bibinfo  {journal} {Physical Review Letters}\ }\textbf {\bibinfo {volume} {12}},\ \bibinfo {pages} {592–595} (\bibinfo {year} {1964})}\BibitemShut {NoStop}%
\bibitem [{\citenamefont {Ballmann}\ \emph {et~al.}(2015)\citenamefont {Ballmann}, \citenamefont {Thompson}, \citenamefont {Traverso}, \citenamefont {Meng}, \citenamefont {Scully},\ and\ \citenamefont {Yakovlev}}]{Ballmann2015}%
  \BibitemOpen
  \bibfield  {author} {\bibinfo {author} {\bibfnamefont {C.~W.}\ \bibnamefont {Ballmann}}, \bibinfo {author} {\bibfnamefont {J.~V.}\ \bibnamefont {Thompson}}, \bibinfo {author} {\bibfnamefont {A.~J.}\ \bibnamefont {Traverso}}, \bibinfo {author} {\bibfnamefont {Z.}~\bibnamefont {Meng}}, \bibinfo {author} {\bibfnamefont {M.~O.}\ \bibnamefont {Scully}}, \ and\ \bibinfo {author} {\bibfnamefont {V.~V.}\ \bibnamefont {Yakovlev}},\ }\bibfield  {title} {\enquote {\bibinfo {title} {Stimulated {Brillouin} scattering microscopic imaging},}\ }\href {\doibase 10.1038/srep18139} {\bibfield  {journal} {\bibinfo  {journal} {Scientific Reports}\ }\textbf {\bibinfo {volume} {5}} (\bibinfo {year} {2015}),\ 10.1038/srep18139}\BibitemShut {NoStop}%
\bibitem [{\citenamefont {Al-Dabbagh}\ and\ \citenamefont {Al-Raweshidy}(2016)}]{AlDabbagh2016}%
  \BibitemOpen
  \bibfield  {author} {\bibinfo {author} {\bibfnamefont {R.}~\bibnamefont {Al-Dabbagh}}\ and\ \bibinfo {author} {\bibfnamefont {H.}~\bibnamefont {Al-Raweshidy}},\ }\bibfield  {title} {\enquote {\bibinfo {title} {Photonic methods of millimeter-wave generation based on {Brillouin} fiber laser},}\ }\href {\doibase 10.1016/j.optlastec.2015.12.005} {\bibfield  {journal} {\bibinfo  {journal} {Optics and Laser Technology}\ }\textbf {\bibinfo {volume} {79}},\ \bibinfo {pages} {124–131} (\bibinfo {year} {2016})}\BibitemShut {NoStop}%
\bibitem [{\citenamefont {Corredera}\ \emph {et~al.}(2013)\citenamefont {Corredera}, \citenamefont {Galindo-Santos}, \citenamefont {Senent}, \citenamefont {Prieto}, \citenamefont {Carrasco-Sanz},\ and\ \citenamefont {Mart{\'i}n-L{\'o}pez}}]{Corredera2013}%
  \BibitemOpen
  \bibfield  {author} {\bibinfo {author} {\bibfnamefont {P.}~\bibnamefont {Corredera}}, \bibinfo {author} {\bibfnamefont {J.}~\bibnamefont {Galindo-Santos}}, \bibinfo {author} {\bibfnamefont {F.~D.}\ \bibnamefont {Senent}}, \bibinfo {author} {\bibfnamefont {F.}~\bibnamefont {Prieto}}, \bibinfo {author} {\bibfnamefont {A.}~\bibnamefont {Carrasco-Sanz}}, \ and\ \bibinfo {author} {\bibfnamefont {S.}~\bibnamefont {Mart{\'i}n-L{\'o}pez}},\ }\bibfield  {title} {\enquote {\bibinfo {title} {{Applications of a femtocomb laser optically filtered by stimulated Brillouin scattering in an optical fiber}},}\ }in\ \href {\doibase 10.1117/12.2041811} {\emph {\bibinfo {booktitle} {Photonics North 2013}}},\ Vol.\ \bibinfo {volume} {8915},\ \bibinfo {editor} {edited by\ \bibinfo {editor} {\bibfnamefont {P.}~\bibnamefont {Cheben}}, \bibinfo {editor} {\bibfnamefont {J.}~\bibnamefont {Schmid}}, \bibinfo {editor} {\bibfnamefont {C.}~\bibnamefont {Boudoux}}, \bibinfo {editor} {\bibfnamefont {L.~R.}\ \bibnamefont {Chen}}, \bibinfo
  {editor} {\bibfnamefont {A.}~\bibnamefont {Del{\^a}ge}}, \bibinfo {editor} {\bibfnamefont {S.}~\bibnamefont {Janz}}, \bibinfo {editor} {\bibfnamefont {R.}~\bibnamefont {Kashyap}}, \bibinfo {editor} {\bibfnamefont {D.~J.}\ \bibnamefont {Lockwood}}, \bibinfo {editor} {\bibfnamefont {H.-P.}\ \bibnamefont {Loock}}, \ and\ \bibinfo {editor} {\bibfnamefont {Z.}~\bibnamefont {Mi}}},\ \bibinfo {organization} {International Society for Optics and Photonics}\ (\bibinfo  {publisher} {SPIE},\ \bibinfo {year} {2013})\ p.\ \bibinfo {pages} {89151B}\BibitemShut {NoStop}%
\bibitem [{\citenamefont {Murphy}\ \emph {et~al.}(2022)\citenamefont {Murphy}, \citenamefont {Yerolatsitis}, \citenamefont {Birks},\ and\ \citenamefont {Stone}}]{Murphy2022}%
  \BibitemOpen
  \bibfield  {author} {\bibinfo {author} {\bibfnamefont {L.~R.}\ \bibnamefont {Murphy}}, \bibinfo {author} {\bibfnamefont {S.}~\bibnamefont {Yerolatsitis}}, \bibinfo {author} {\bibfnamefont {T.~A.}\ \bibnamefont {Birks}}, \ and\ \bibinfo {author} {\bibfnamefont {J.~M.}\ \bibnamefont {Stone}},\ }\bibfield  {title} {\enquote {\bibinfo {title} {Stack, seal, evacuate, draw: a method for drawing hollow-core fiber stacks under positive and negative pressure},}\ }\href {\doibase 10.1364/oe.470599} {\bibfield  {journal} {\bibinfo  {journal} {Optics Express}\ }\textbf {\bibinfo {volume} {30}},\ \bibinfo {pages} {37303} (\bibinfo {year} {2022})}\BibitemShut {NoStop}%
\bibitem [{\citenamefont {Litchinitser}\ \emph {et~al.}(2002)\citenamefont {Litchinitser}, \citenamefont {Abeeluck}, \citenamefont {Headley},\ and\ \citenamefont {Eggleton}}]{Litchinitser2002}%
  \BibitemOpen
  \bibfield  {author} {\bibinfo {author} {\bibfnamefont {N.~M.}\ \bibnamefont {Litchinitser}}, \bibinfo {author} {\bibfnamefont {A.~K.}\ \bibnamefont {Abeeluck}}, \bibinfo {author} {\bibfnamefont {C.}~\bibnamefont {Headley}}, \ and\ \bibinfo {author} {\bibfnamefont {B.~J.}\ \bibnamefont {Eggleton}},\ }\bibfield  {title} {\enquote {\bibinfo {title} {Antiresonant reflecting photonic crystal optical waveguides},}\ }\href {\doibase 10.1364/OL.27.001592} {\bibfield  {journal} {\bibinfo  {journal} {Opt. Lett.}\ }\textbf {\bibinfo {volume} {27}},\ \bibinfo {pages} {1592--1594} (\bibinfo {year} {2002})}\BibitemShut {NoStop}%
\bibitem [{\citenamefont {Uebel}\ \emph {et~al.}(2016)\citenamefont {Uebel}, \citenamefont {G\"{u}nendi}, \citenamefont {Frosz}, \citenamefont {Ahmed}, \citenamefont {Edavalath}, \citenamefont {Ménard},\ and\ \citenamefont {Russell}}]{Uebel2016}%
  \BibitemOpen
  \bibfield  {author} {\bibinfo {author} {\bibfnamefont {P.}~\bibnamefont {Uebel}}, \bibinfo {author} {\bibfnamefont {M.~C.}\ \bibnamefont {G\"{u}nendi}}, \bibinfo {author} {\bibfnamefont {M.~H.}\ \bibnamefont {Frosz}}, \bibinfo {author} {\bibfnamefont {G.}~\bibnamefont {Ahmed}}, \bibinfo {author} {\bibfnamefont {N.~N.}\ \bibnamefont {Edavalath}}, \bibinfo {author} {\bibfnamefont {J.-M.}\ \bibnamefont {Ménard}}, \ and\ \bibinfo {author} {\bibfnamefont {P.~S.}\ \bibnamefont {Russell}},\ }\bibfield  {title} {\enquote {\bibinfo {title} {{Broadband robustly single-mode hollow-core PCF by resonant filtering of higher-order modes}},}\ }\href {\doibase 10.1364/ol.41.001961} {\bibfield  {journal} {\bibinfo  {journal} {Optics Letters}\ }\textbf {\bibinfo {volume} {41}},\ \bibinfo {pages} {1961} (\bibinfo {year} {2016})}\BibitemShut {NoStop}%
\bibitem [{Her()}]{HeraeusWeb}%
  \BibitemOpen
  \href@noop {} {}\bibinfo {howpublished} {\url{https://www.heraeus-conamic.com/}}\BibitemShut {NoStop}%
\bibitem [{\citenamefont {Jackson}\ \emph {et~al.}(2024)\citenamefont {Jackson}, \citenamefont {Jasion}, \citenamefont {Bradley}, \citenamefont {Poletti},\ and\ \citenamefont {Davidson}}]{Jackson2024}%
  \BibitemOpen
  \bibfield  {author} {\bibinfo {author} {\bibfnamefont {G.}~\bibnamefont {Jackson}}, \bibinfo {author} {\bibfnamefont {G.~T.}\ \bibnamefont {Jasion}}, \bibinfo {author} {\bibfnamefont {T.~D.}\ \bibnamefont {Bradley}}, \bibinfo {author} {\bibfnamefont {F.}~\bibnamefont {Poletti}}, \ and\ \bibinfo {author} {\bibfnamefont {I.~A.}\ \bibnamefont {Davidson}},\ }\bibfield  {title} {\enquote {\bibinfo {title} {{Three stage HCF fabrication technique for high yield, broadband UV-visible fibers}},}\ }\href {\doibase 10.1364/oe.507703} {\bibfield  {journal} {\bibinfo  {journal} {Optics Express}\ }\textbf {\bibinfo {volume} {32}},\ \bibinfo {pages} {7720} (\bibinfo {year} {2024})}\BibitemShut {NoStop}%
\bibitem [{\citenamefont {Harrington}\ \emph {et~al.}(2024)\citenamefont {Harrington}, \citenamefont {Mears}, \citenamefont {Stone}, \citenamefont {Wadsworth}, \citenamefont {Knight},\ and\ \citenamefont {Birks}}]{Harrington2024a}%
  \BibitemOpen
  \bibfield  {author} {\bibinfo {author} {\bibfnamefont {K.}~\bibnamefont {Harrington}}, \bibinfo {author} {\bibfnamefont {R.}~\bibnamefont {Mears}}, \bibinfo {author} {\bibfnamefont {J.~M.}\ \bibnamefont {Stone}}, \bibinfo {author} {\bibfnamefont {W.~J.}\ \bibnamefont {Wadsworth}}, \bibinfo {author} {\bibfnamefont {J.~C.}\ \bibnamefont {Knight}}, \ and\ \bibinfo {author} {\bibfnamefont {T.~A.}\ \bibnamefont {Birks}},\ }\bibfield  {title} {\enquote {\bibinfo {title} {Optical absorption spectrum reveals gaseous chlorine in anti-resonant hollow core fibers},}\ }\href {\doibase 10.1364/oe.537473} {\bibfield  {journal} {\bibinfo  {journal} {Optics Express}\ }\textbf {\bibinfo {volume} {32}},\ \bibinfo {pages} {38072} (\bibinfo {year} {2024})}\BibitemShut {NoStop}%
\bibitem [{\citenamefont {Hillebrands}(1999)}]{Hillebrands1999}%
  \BibitemOpen
  \bibfield  {author} {\bibinfo {author} {\bibfnamefont {B.}~\bibnamefont {Hillebrands}},\ }\bibfield  {title} {\enquote {\bibinfo {title} {Progress in multipass tandem {Fabry--Perot} interferometry: {I.} {A} fully automated, easy to use, self-aligning spectrometer with increased stability and flexibility},}\ }\href {\doibase 10.1063/1.1149637} {\bibfield  {journal} {\bibinfo  {journal} {Rev. Sci. Instrum.}\ }\textbf {\bibinfo {volume} {70}},\ \bibinfo {pages} {1589--1598} (\bibinfo {year} {1999})}\BibitemShut {NoStop}%
\bibitem [{\citenamefont {Kargar}\ and\ \citenamefont {Balandin}(2021)}]{Kargar2021}%
  \BibitemOpen
  \bibfield  {author} {\bibinfo {author} {\bibfnamefont {F.}~\bibnamefont {Kargar}}\ and\ \bibinfo {author} {\bibfnamefont {A.~A.}\ \bibnamefont {Balandin}},\ }\bibfield  {title} {\enquote {\bibinfo {title} {Advances in {B}rillouin--{M}andelstam light-scattering spectroscopy},}\ }\href {\doibase 10.1038/s41566-021-00836-5} {\bibfield  {journal} {\bibinfo  {journal} {Nat. Photonics}\ }\textbf {\bibinfo {volume} {15}},\ \bibinfo {pages} {720--731} (\bibinfo {year} {2021})}\BibitemShut {NoStop}%
\bibitem [{\citenamefont {Lindsay}, \citenamefont {Anderson},\ and\ \citenamefont {Sandercock}(1981)}]{Lindsay1981}%
  \BibitemOpen
  \bibfield  {author} {\bibinfo {author} {\bibfnamefont {S.~M.}\ \bibnamefont {Lindsay}}, \bibinfo {author} {\bibfnamefont {M.~W.}\ \bibnamefont {Anderson}}, \ and\ \bibinfo {author} {\bibfnamefont {J.~R.}\ \bibnamefont {Sandercock}},\ }\bibfield  {title} {\enquote {\bibinfo {title} {Construction and alignment of a high performance multipass vernier tandem {Fabry}--{P\'{e}rot} interferometer},}\ }\href {\doibase 10.1063/1.1136479} {\bibfield  {journal} {\bibinfo  {journal} {Rev. Sci. Instrum.}\ }\textbf {\bibinfo {volume} {52}},\ \bibinfo {pages} {1478--1486} (\bibinfo {year} {1981})}\BibitemShut {NoStop}%
\bibitem [{\citenamefont {Mock}, \citenamefont {Hillebrands},\ and\ \citenamefont {Sandercock}(1987)}]{Mock1987}%
  \BibitemOpen
  \bibfield  {author} {\bibinfo {author} {\bibfnamefont {R.}~\bibnamefont {Mock}}, \bibinfo {author} {\bibfnamefont {B.}~\bibnamefont {Hillebrands}}, \ and\ \bibinfo {author} {\bibfnamefont {R.}~\bibnamefont {Sandercock}},\ }\bibfield  {title} {\enquote {\bibinfo {title} {Construction and performance of a {Brillouin} scattering set-up using a triple-pass tandem {Fabry--Perot} interferometer},}\ }\href {\doibase 10.1088/0022-3735/20/6/017} {\bibfield  {journal} {\bibinfo  {journal} {J. Phys. E.}\ }\textbf {\bibinfo {volume} {20}},\ \bibinfo {pages} {656--659} (\bibinfo {year} {1987})}\BibitemShut {NoStop}%
\end{thebibliography}%

\end{document}